\documentclass[11pt]{article}
\usepackage{geometry}
\usepackage{graphicx}
\usepackage{amsmath}
\renewcommand{\vec}[1]{\mathbf{#1}}
\newcommand{\sfrac}[2]{\mbox{$\frac{#1}{#2}$}}
\newcommand{\trace}{\mathrm{Tr}}
\newcommand{\half}{\mbox{$\frac{1}{2}$}}
\newcommand{\fuvc}[1]{\hat{#1}}
\newcommand{\ltvc}[1]{{\left\{\fuvc{#1}\right\}}}
\newcommand{\ltvac}{{\left\{0\right\}}}
\newcommand{\irrep}[1]{{\chi_{#1}}}
\newcommand{\ket}[1]{\left|#1\right\rangle}
\newcommand{\bra}[1]{\left\langle#1\right|}
\newcommand{\braket}[2]{\left\langle #1|#2 \right\rangle}
\newcommand{\be}{\begin{equation}}
\newcommand{\ee}{\end{equation}}
\newcommand{\bea}{\begin{eqnarray}}
\newcommand{\eea}{\end{eqnarray}}

\begin{document}
\title{A Coupled-Cluster Formulation of Hamiltonian Lattice Field Theory:
       The Non-Linear Sigma Model}
\author{N.E. Ligterink\thanks{e-mail: N.E.Ligterink@umist.ac.uk}, N.R. Walet\thanks{e-mail: N.R.Walet@umist.ac.uk}, and R.F. Bishop\thanks{e-mail: R.F.Bishop@umist.ac.uk} \\
Department of Physics, UMIST, PO Box 88, Manchester, M60 1QD,  UK}
\maketitle
\thispagestyle{myheadings}

\begin{abstract}
We apply the coupled cluster method (CCM) to the Hamiltonian 
version of the latticised $O(4)$ non-linear sigma model. 
The method, which was initially developed for the accurate description
of quantum many-body systems, gives rise to two distinct approximation
schemes. These approaches are compared with each other as well as with 
some other Hamiltonian approaches. Our study of both the ground state and 
collective excitations leads to indications of a possible chiral phase 
transition as the lattice spacing is varied.
\end{abstract}

\newpage

\vspace{5mm}
\noindent
{\Large \bf Contents} \\
\vspace{3mm}

\noindent
{1} Introduction;
{2} The $O(4)$ non-linear sigma model;
{3} Elements of coupled cluster theory;
{3.1} The standard form of the CCM;
{3.2} The functional form of the CCM;
{4} The operatorial form of the CCM for classical spin models;
{4.1} The LSUB2 approximation;
{4.2} The SUB2-$n$ approximations;
{4.3} Full SUB2 calculations;
{4.4} Collective excitations;
{5} The functional form of the CCM for classical spin models;
{5.1} The LSUB2 aproximation;
{5.2} The SUB2-$n$-$m$ approximations;
{5.3} Collective excitations;
{6} Solution and Results;
{6.1} Numerical methods;
{6.2} LSUB2 Results;
{6.3} SUB2-$n$(-$m$) Results;
{6.4} Excitations;
{6.5} Comparison with other work;
{7} Conclusions and outlook;
{A} Quantisation;
{B} Gegenbauer polynomials

\section{Introduction}\label{sec:intro}

The physics of particles interacting through the strong forces at
low energies is dominated by pions
(see, e.g., the discussion in Ref. \cite{DGH92}). In the hadronic 
sector the exchange of single
and correlated pions between nucleons determines the long-range and
intermediate-range behaviour of the nuclear force. 
The QCD Lagrangian, which is supposed to govern all strong interaction
physics, has a chiral symmetry. Were this symmetry realised in nature,
we would expect all hadrons to come in chiral pairs in which each hadron
has a partner of the same mass but opposite parity. The fact that this 
does not actually  occur
is due to the effect of spontaneous breaking of the chiral
symmetry which arises from the development of a condensate of quarks 
and gluons. The pions are interpreted as the Goldstone bosons which
occur in such a scenario. If QCD were exactly chirally symmetric we
would expect to find them at zero mass. The fact that their rest-mass
is about 140 MeV/c$^{2}$ is due to the effect of the small masses carried by
the quarks. This can be expressed in the Gell-Mann-Oakes-Renner
relation \cite{GOR}, which relates the pion mass to the quark condensate 
and to the up and down quark masses, $m_u$ and $m_d$ respectively,
\begin{equation}
m_\pi^2 = -\left( \frac{2}{f_\pi} \right)
\langle \bar \psi \psi \rangle (m_u+m_d).  
\end{equation}
Here $\langle \bar \psi \psi \rangle$ is the quark condensate in vacuum. 

In recent years a simple scaling argument
by Brown and Rho \cite{BrownRho} has sparked
heated discussions about how mesons behave in a medium, where the quark
condensate changes due to a partial restoration of chiral symmetry.
The same
question can be asked for finite temperature. One believes that for
temperatures of around the order of the pion mass the chiral
condensate will disappear and chiral symmetry will be restored
\cite{BW94}. It is believed that in heavy ion collisions it is 
possible to inject enough energy to create such a phase.
It is not yet clear whether such a plasma can reach
thermal equilibrium, nor whether the time scale is long enough for the
chiral phase to develop.

All these issues have renewed interest in the dynamics of systems of pions.  At
the same time a revival has occurred in so-called baryon chiral
perturbation theory \cite{BchPt}, where the most general effective
Lagrangian compatible with chiral symmetry is constructed, and all
constants are fitted to experiments. This approach, which has met
great success in the meson sector \cite{DGH92}, is now
actively being pursued in the baryonic sector.

As mentioned above, pions find their origin in the chiral symmetry,
which can be related to the fundamental QCD Lagrangian. In the
dynamical breaking of the symmetry the chiral partners of hadrons
disappear and a set of three massless bosons, the Goldstone bosons,
emerges.  Although the mechanism is well understood, it is difficult to
see explicitly how pions arise from the original QCD Lagrangian,
especially since the pions are known to be composite particles.
Therefore one often works with an effective Lagrangian which is
compatible with the consequences of spontaneous symmetry breaking. 
The pion dynamics is somewhat restricted, since we have partial conservation
of the axial current (PCAC), which limits the interactions we can 
build into our models. One model that satisfies these constraints
 is the $O(4)$ non-linear sigma
model, which we shall study in this paper.

The class of non-linear sigma models plays an illustrative  r\^ole in many
physical phenomena \cite{PW83,PW84}. In particular, the $O(4)$ version
of the model can be used to describe the dynamics of pions. In that case
the $O(4)$ symmetry can be broken explicitly, but the isospin $O(3)$
sub-symmetry must be retained. Using a perturbative approach for
the model in 2 dimensions (1 space and 1 time) one finds that it
exhibits both infrared and ultraviolet divergences \cite{PW83}. The latter are
thought to be responsible for dimensional transmutation, the process
whereby the dimensionless coupling constant acquires a dependence on a
fundamental length, such as the QCD scale in QCD.

The short-distance behaviour has to be regularised, which is generally
done by putting the model onto a lattice. For some finite lattice spacing
one expects a phase transition between a system with essentially free
rotators which interact weakly (at large lattice spacings), and a
system of tightly bound rotators which exhibit only collective behaviour
(at small lattice spacings). In the former phase all excitations are
massive and are characterised by the quantum numbers of the free rotator. In
the latter case the excitation energy gap disappears. However, for two
dimensions (one space and one time) the situation is different. From
exact results it is known that the phase transition disappears due
to strong interactions among the Goldstone bosons \cite{PW83}. The
excitations remain massive, although their masses decrease with
decreasing lattice spacing. This result is confirmed in dimensional
regularisation \cite{BZG76}, as is the asymptotic freedom of 
the two-dimensional  non-linear sigma model.

Most results in $O(4)$ theory have been derived in a Euclidean framework
\cite{BC96}. In that case there is also a clear correspondence with  
high-temperature spin models \cite{Sta68}. In a Euclidean framework space
and time are treated on an equal footing. Not only is the time coordinate
discretised but it is also taken to be imaginary. Apart from 
the ground-state energy
it is very hard to recover properties of the system, and it is not clear
whether the Euclidean and the Minkowskian (or Hamiltonian) methods yield
the same result in a non-perturbative framework, especially since the
lattice regularisation is implemented in such different ways in 
the two methods.

In a Hamiltonian framework time is taken to be real and continuous (as it is
in nature). One can construct quantum-mechanical (Schr\"odinger) states 
and calculate basic expectation
values. However, for gauge theories the gauge has to be fixed when 
constructing a Hamiltonian, 
and therefore gauge invariance might be destroyed in subsequent
approximations. Space is discretised on a lattice, and the lattice spacing
is the only dimensional parameter in the model. The phase transition
from the disoriented (symmetric) phase of free rotators to the oriented 
(spontaneously broken symmetry) phase of tightly
bound rotators can be seen as the dimensional transmutation, which sets 
the scale
at which the system starts to exhibit the basic continuum behaviour.

As the non-linear sigma model is an effective field theory for pions
these features should put pion dynamics in a broader context. Indeed,
we see that in low-energy pion dynamics the interaction vanishes
as the momenta go to zero, which follows rather naturally from the fact that
the pions are the Goldstone bosons of the chiral symmetry and the
interaction is dictated by the spontaneous symmetry-breaking mechanism.
However, since the pion is a composite particle the connection should
break down for higher momenta (preferably somewhere near the QCD scale).
This makes it difficult to connect excitations in the large lattice spacing
systems we are investigating to actual physical particles. The system
is more closely related to the dynamics of rotators described above.

Apart from its intrinsic physical interest the non-linear sigma model 
is an ideal
testing ground for non-perturbative methods. Although the Lagrangian is
fairly simple and the degrees of freedom are restricted, it still has the
richness of a full physical theory, whose properties should be correctly described by
a proper non-perturbative method for the study of lattice field
theories.  
The structure of field theories on the lattice is such that it
begs to be treated by many-body techniques which first allow
the ground state to be approximated in a well-determined and controlled way.
Recently, Chin \cite{Shi97} has performed such a  many-body calculation 
for the model being studied in this paper, using a variational technique 
embedded in a more general  correlated basis function (CBF) approach. 
By contrast with the techniques used here,
the CBF method, which uses a generalized Jastrow-correlated wave
function cannot easily be formulated as a set of algebraic equations.
Instead, it needs to be solved by Monte Carlo techniques on a finite portion
of the infinite lattice, just as in most lattice-gauge approaches. The limit 
to an infinite lattice needs to be taken at the end of the calculation, 
using heuristic or theoretical scaling arguments.

Our approach to the problem is the so-called coupled cluster method (CCM).
The CCM is now widely acknowledged to provide one of the
most widely applicable and most powerful of all microscopic formulations
of quantum many-body theory. In the many applications which have  so far 
been made in 
such diverse fields as nuclear physics, quantum field theory, condensed matter
physics, quantum magnetism, and quantum chemistry \cite{BishopChim},
 the CCM has been found to
provide results which are among the most accurate available. Indeed, it has 
been shown to give results comparable to those from large-scale quantum Monte Carlo
calculations in those cases where the latter can be performed. The interested
reader is referred to Ref.~\cite{BishopChim} for a survey of the method and an overview of its applications.

The CCM is formulated in terms of cluster correlation functions. These 
correlation
functions, although typically only correlating the particles (or spins, or fields) on
clusters of lattice sites 
with a finite spatial extent, do nevertheless
extend the correlated wave function over the infinite lattice. In the CCM, unlike in CBF
and most other lattice calculations, the lattice may be treated as infinite
from the outset, as we will see below.

In particular, the CCM has recently been applied to simple 
$U(1)$ lattice gauge field theory
\cite{BBD96,BBD97}, and has been shown to provide an interesting scheme for
studying the properties of such models. In the current work we introduce
 a variant of
the CCM that works with functions rather than with the more usual
operators.
This approach will be discussed in considerable detail in the present paper,
although we shall succeed in constructing an operatorial approach as well.

The remainder of this paper is organised as follows. 
In Sec.~\ref{sec:O4} we briefly discuss the $O(4)$ non-linear sigma model
and the Hamiltonian we construct from it (with some details relegated to
Appendix~\ref{app:Quant}). Then, in Sec.~\ref{sec:CCM}, we shall
succinctly introduce the traditional form of the CCM and, in some more detail,
the functional form. In Sec.~\ref{sec:oper} we lay the
theoretical framework for an investigation of the $O(4)$ non-linear
sigma model using the traditional CCM approach. The same is done in 
Sec.~\ref{sec:func} for the functional method. In Sec.~\ref{sec:SolRes}
we collect the results from the various schemes, and contrast these both
among themselves and with the few other results available.
Finally we give a summary and outlook in Sec.~\ref{sec:Conc}.

\section{The $O(4)$ non-linear sigma model}\label{sec:O4}

The $O(4)$ non-linear sigma model is defined by the Lagrangian density
\begin{equation} 
{\mathcal L} = \frac{1}{4} \trace(\partial_\mu U \partial^\mu U^\dagger),
\end{equation}
 where $U$ is an $SU(2)$ matrix-valued field of $2\times2$ unitary matrices,
which is often conveniently represented by a sum of unit and Pauli
matrices,
\begin{equation} 
U = n_4 I + i \vec \tau \cdot \vec n,
\end{equation} 
and the unitarity constrains the four-vector $\fuvc n=(\vec n, n_4)$ 
to have unit length.

We shall study the model on a spatial grid  which is a cubic lattice 
in three dimensions
and with continuous time,
in a Hamiltonian formalism. We replace the spatial derivatives by 
finite differences, with a lattice spacing $a$.
Since the time-derivative part of the
Lagrangian describes a freely rotating four-vector constrained to the
unit sphere, it is not implausible that the kinetic energy 
in the Hamiltonian form at each point in
space can be reduced to the angular part of the Laplacian, as
is shown in Appendix~\ref{app:Quant}. Another way to represent this result is
\begin{eqnarray} 
H &=&
\half \sum_\vec{i} \vec{I}_{\vec{i}}^2 + 
\frac{1}{4a^2} \sum_{\langle \vec{ i}\vec{j} \rangle}
\trace( (U_{\vec{i}}-U_{\vec{j}})(U^\dagger_{\vec{i}}-U^\dagger_{\vec{j}}))
\nonumber\\ &=&
\half \sum_{\vec{i}} \vec{I}_{\vec{i}}^2 + 
\frac{1}{a^2} \sum_{\langle \vec{ i}\vec{j} \rangle}
\left[1-\cos(\theta_{\vec{i}\vec{j}})\right].
\end{eqnarray} 
The lattice Hamiltonian contains a trivial factor $a^D$:  the volume of the
unit lattice cell in $D$ spatial dimensions. We implicitly assume that
all expectation values scale with this volume.
Here $I_a$ can be taken as either the left or right SU(2) generators,
\begin{equation} [{I}_{a\vec{i}}, U_{\vec{j}}] = i\tau_{a} 
U_{\vec{i}}\delta_{\vec{i}\vec{j}}
\qquad{\rm or}\qquad[{I}_{a\vec{i}}, U_{\vec{j}}] =
i U_{\vec{i}}\delta_{\vec{i}\vec{j}}\tau_{a} ,
\end{equation} 
and 
$\theta_{\vec{i}\vec{j}} \equiv \fuvc{n}_{\vec{i}} \cdot \fuvc{n}_{\vec{j}}$ 
is the angle between the two unit vectors
$\fuvc{n}_{\vec{i}}$ and $ \fuvc{n}_{\vec{j}}$ describing $U_{\vec{i}}$ and $U_{\vec{j}}$.
We use here the convention related to the $O(4)$ symmetry. Factors of a half might
appear in other conventions related to the
isospin symmetry (and see, e.g., Chin~\cite{Shi97}). We note that these
different conventions will also lead to  corresponding alternative definitions
of the kinetic energy. We ourselves choose the convention which yields the kinetic 
energy of a unit mass on the four-dimensional unit sphere.
Relabelling $\lambda=\frac{1}{a^2}$, we have the classical spin
model
\begin{equation}
 H =
\half \sum_{\vec{i}} \vec{I}_{\vec{i}}^2 + 
\lambda \sum_{\langle \vec{ i}\vec{j} \rangle}(
1-\cos \theta_{\vec{i}\vec{j}} ),
\label{fullhamil}
\end{equation} 
where the sum over $\langle \vec{i}\vec{j} \rangle$ runs over all 
nearest-neighbour pairs, and counts each pair (or lattice link) only once.

The Hamiltonian can thus be interpreted as that of freely rotating 
unit vectors (or quantum rotors), with a nearest-neighbour coupling that tends to align 
them. It is this competition between free rotation and an alignment
force that makes these models 
interesting to study. It also shows that they are identical to
classical spin systems on a lattice, which can be obtained as 
either the large-$J$ limit of a finite-$J$ lattice spin model, or
in the calculation of the partition function of a {\em classical}
lattice spin problem.

\section{Elements of coupled cluster theory}\label{sec:CCM}

We start with an $O(N)$ Hamiltonian of the form
\begin{equation}H = \sum_{\vec{i}} K_{\vec{i}} +
\lambda\sum_{\langle \vec{i}\vec{j} \rangle} (1-\fuvc n_{\vec{i}}\cdot
\fuvc n_{\vec{j}}) . \label{hamil}
\end{equation} Here $K_{\vec{i}}$ is the kinetic energy on the $N$-dimensional unit
sphere at lattice point $\vec{i}$.  The eigenstates of $K_{\vec{i}}$ are the 
(hyper)spherical harmonics
$\ket{J \alpha}_{\vec{i}}$,
\begin{equation}
K_\vec{i} \ket{J \alpha}_{\vec{i}} = \epsilon_J \ket{J \alpha}_{\vec{i}}.
\end{equation} 
Here we have divided the quantum numbers into those $(J)$ that determine the
eigenvalues of $K_{\vec{i}}$, and those $(\alpha)$ that label the different degenerate eigenstates.

There are now at least two distinct ways to formulate a coupled-cluster 
approach for such a
classical spin model. Firstly, there is an operatorial approach, which
models itself 
closely on the standard formulation of the CCM, albeit with rather
artificial operators. Secondly, there also exists  a functional approach,
 which was first
applied to lattice QED problems. Although the latter functional approach
appears to be more natural for the present study,
both methods are investigated and compared below.

\subsection{The standard form of the CCM}

Coupled cluster theory is one of the mainstays of modern microscopic 
quantum many-body
theory, and reviews of the method are available (and see, e.g., 
ref.~\cite{BishopChim}). We shall therefore only give a very short summary of the
salient points of the method; for full details see Ref.~\cite{BishopChim}
and references cited therein.

The key feature of coupled cluster theory is its
 construction of a
systematically improvable approximation to the exact correlated ground state
of an interacting system, by the action of  the exponential of an operator
$\hat S$  on some suitable model state or bare vacuum,
$| 0 \rangle$, which may (but does not have to) be chosen as the exact ground 
state in some limit.
 No attempt is made to normalise
this state against itself, but one rather constructs a bra state in a bi-orthogonal
system such that the inner product of the bra and the ket states remains
unity. One starts from a set of (generalized)  multi-particle or
multi-mode creation operators
$a^\dagger_n$, where the corresponding hermitian conjugate annihilation 
operators destroy
the vacuum, $a_n | 0 \rangle =0 $. We now write the correlated ground state as
\begin{equation}
\ket{\Psi} = \exp\left(\sum_n S_n a^\dagger_n\right) \ket{0} .
\label{eqPsi}
\end{equation}
This state satisfies an intermediate normalisation condition,
$\braket{0}{\Psi}=1$.  In many-body applications, the exponential
Ansatz of Eq.~(\ref{eqPsi}) guarantees proper size-extensivity and conformity
with the Goldstone linked-cluster theorem even when approximations
are made. By contrast, the corresponding linear parametrisation which 
characterises the  corresponding configuration-interaction method 
(or generalized many-body shell model) will not, in general, 
obey either of these important properties when approximations are made.
One also parameterises the corresponding correlated bra state as
\be
\bra{\smash[t]{\tilde \Psi}} = \bra{0}\left(1+\sum_n \tilde S_n
a_n\right)\exp\left(-\sum_n S_n a^\dagger_n\right) ,
\label{eqBra}
\end{equation}
where the coefficients $S_n$ and $\tilde S_n$ are to be determined independently
by an extremum requirement, which can be given the variational form
\be
\delta (I[S,\tilde S])=0,
\end{equation} with
\begin{equation}
I[S,\tilde S] = \bra{\smash[t]{\tilde\Psi}} H \ket{\Psi} ,
\end{equation}
where $I[S, \tilde S]$  is the functional with respect to the infinite set
of coefficients $S=\{S_n\ |\  n = 1,2, \cdots \}$ and 
$\tilde S=\{\tilde S_n\ |\  n = 1,2, \cdots \}$.
In practice, however, we need to truncate the sum over creation and annihilation
operators, and a consistent truncation scheme is found by restricting the two
sets of coefficients in the same manner.
For this choice we also find that the Hellmann-Feynman
theorem is satisfied, i.e.
\be
\bra{\smash[t]{\tilde\Psi}} \partial_\lambda H \ket{\Psi} = \partial_\lambda E,
\end{equation}
where $E$ is the ground-state energy, $E\equiv \bra{\smash[t]{\tilde\Psi}} H \ket{\Psi}$.
Finally, in a  formulation of the theory where the coefficients
$S_n$ and $\tilde S_n$ are time-dependent, we find that $S_n$ and
$\tilde S_n$ are canonically conjugate variables. There is a deep connection
between these three properties, and we cannot give up one without giving
up something else as well ~\cite{BishopChim}. In particular we note that
although in untruncated, and hence exact, parametrisations of the form
of Eqs.~(\ref{eqPsi}) and (\ref{eqBra}), the hermiticity relation $\tilde {\ket{ \Psi}}  = \ket{\Psi} / \langle \Psi | \Psi \rangle $ would hold, when
Eqs.~(\ref{eqPsi}) and (\ref{eqBra}) are truncated this exact hermiticity
is broken in general. Although Arponen \cite{Arp97} has discussed means to 
restore the hermiticity within the CCM formalism, we do not pursue this
point further here.

\subsection{The functional form of the CCM}

The functional form is similar to the method sketched
above, and as far as we are aware has first been used in
refs. \cite{BBD96,BBD97}. The key ingredient is the parametrisation of 
the ground-state ket wave function in the exponential form
\be
\braket{\{r\}}{ \Psi} = \exp\left[S(\{r\})\right],
\end{equation} 
where $\ket{\{r\}}$ is some suitably chosen complete set of wave
functions. Such an approach appears to be naturally suited to lattice
Hamiltonians, where a complete set of functions can often be chosen to be
the product of complete sets at each lattice site.
Truncations, which in the operator form are performed on the
number and type of creation operators, are now performed on the
functional dependence of the function $S(\{r\})$, typically by expansion
in a complete set of appropriate functions. Once we construct
the mode of action of the Hamiltonian on functions of the variables 
${\{r\}}$ -- typically
a differential, but at worst an integro-differential, operator -- we can
proceed as before and use the respective bra-state parametrisation,
\be
\braket{\smash[t]{\tilde \Psi}}{\{r\}} = [1+\tilde S(\{r\})] \exp\left[-S(\{r\})\right] ,
\end{equation} 
and a variational functional $I[S,\tilde S]$ defined by
\be
I[S,\tilde S] = \int d\{r\} d\{r'\}[1+\tilde S(\{r\})] 
\exp\left[-S(\{r\})\right] \bra{\{r\}} H \ket{\{r'\}} 
\exp\left[S(\{r'\})\right].
\label{funcfunc}
\end{equation} 
 One typically excludes
a constant term from $S$ and $\tilde S$ (corresponding to the exclusion 
of the identity operator from the sum over the multi-configurational 
creation operators in the various terms in Eqs.~(\ref{eqPsi}) 
and (\ref{eqBra})). However, we now no longer
have intermediate normalisation. Thus, the constant part of 
$\exp\left(S(\{r\})\right)$ is not one, since an arbitrary power of a 
function orthogonal to the constant function can contain a constant part.
In this case we need to enquire  what happens with the Hellmann-Feynman 
theorem and the canonical
variables in the time-dependent formalism? The key  to the answer lies 
in orthogonality.
While this is provided for by Wick's theorem when using the operator form, 
in the functional form this requires an expansion in orthogonal functions,
relative to (typically) a  product of integrations. 
This will be shown to be straightforwardly implementable
 in the lowest set of approximations (SUB2-$n$, where only $n$ two-body
correlations are retained)
for the model under consideration here.  Since many-body orthogonal polynomials 
are not straightforward to define, more
complicated correlations involving more than pairwise correlations
 do not seem to be easy to formulate in the
functional form of the theory. It may still
be that for certain classes of lattice models  a functional 
form is easier to use, and even though we lose some of the elegance
of the standard CCM, we might nevertheless still be able to proceed 
using this method.

\section{The operatorial form of the CCM for classical spin models}
\label{sec:oper}
Let us first investigate the operatorial CCM approach when applied to
classical spin models of the form given in Eq.~(\ref{hamil}). We use
the fact that the uncorrelated ground-state wave function is given by
the constant function (the product over the grid of the lowest eigenstates
$\ket{0}_\vec{i}$ of the kinetic energy operator $K_\vec{i}$).
We now define a set of excitation operators for the CCM by
\begin{equation}C^\dagger_{\vec{i}J\alpha} = \ket{J\alpha}_\vec{i}\bra{0}_{\vec{i}} 
\ ; \ \  J>0,
\end{equation} 
which are chosen such that their hermitian conjugates annihilate 
the vacuum. The quantum number $J$ labels the energy eigenstates of
a single rotor, and $\alpha$ labels the different states for a given
energy. We shall use the notation
\be
\ket{\ltvc n} = \prod_\vec{i} \ket{\fuvc n_\vec{i}},\;\;\;
\ket{\ltvac} = \prod_{\vec{i}}\ket{0_\vec{i}},
\end{equation}
where the states $\ket{\fuvc n_\vec{i}}$ are the standard coordinate 
representation at lattice site $\vec{i}$.
We shall mainly be concerned with the SUB2 family of approximations,
in which only correlations between pairs of spins are incorporated,
and in which the coupled cluster operator can hence be written as
\be
\hat S_2 \equiv  \sum_{ [\vec{i}\vec{j}] } \hat S_{2\vec{i}\vec{j}}
 = \sum_{ I  [\vec{i}\vec{j} ]}
 s_{I} \left[C^\dagger_{\vec{i}I}\otimes
C^\dagger_{\vec{j}I}\right]^{(0)},  \label{eq:SUB2}
\end{equation}
where $I$ is now the label for the energy eigenstates in the relative angle
between the spins at lattice sites $\vec{i}$ and $\vec{j}$.
The sum over $[\vec{i} \vec{j} ]$ is over {\it all} pairs of lattice sites,
$\vec{i}$ and $\vec{j}$, counting each pair once, by contrast with the
sum $\langle \vec{i}\vec{j} \rangle$ which runs over the nearest-neighbour 
pairs only, and which we introduced previously in Eq.~(\ref{fullhamil}).
We have suppressed the $\alpha$ labels, because these
are implicitly summed over in the scalar coupling. Since the only scalar that
we can construct from the two vectors $\fuvc n_\vec{i}$ and $\fuvc n_\vec{j}$
 is their inner product, we find
\begin{eqnarray}
\bra{\fuvc n_\vec{i}, \fuvc n_\vec{j}} \hat S_{2\vec{i}\vec{j}} 
\ket{0_\vec{i},0_\vec{j}}  & = &  
 \frac{1}{\sqrt{\Omega_N}} S_2(\fuvc n_\vec{i}\cdot \fuvc n_\vec{j}) =
 \frac{1}{\sqrt{\Omega_N}} S_2\left(\cos \theta_{\vec{i}\vec{j}}\right) ; \\
\hat S_{2\vec{i}\vec{j}}  & = & \int d \fuvc n_\vec{i} d \fuvc n_\vec{j}
 |{\fuvc n_\vec{i},\fuvc n_\vec{j}} \rangle \, 
 \frac{1}{\sqrt{\Omega_N}} S_2(\cos \theta_{\vec{i}\vec{j}} ) \, \langle {0_\vec{i},0_\vec{j}} | ,
\end{eqnarray} 
where $\Omega_N$ is the surface of the $N$-dimensional unit sphere.
The potential can be expressed in the same way. Since the potential only
depends on the relative angle between nearest-neighbour spins we require
only the scalar coupling of spins
\begin{equation}
 \hat V_{\vec{i}\vec{j}}    =  \int d \fuvc n_\vec{i} d \fuvc n_\vec{j}
 |{\fuvc n_\vec{i},\fuvc n_\vec{j}} \rangle \, \lambda \,  
 (1 - \cos \theta_{\vec{i}\vec{j}} ) \, \langle {\fuvc n_\vec{i},\fuvc n_\vec{j}} | .
\end{equation}

We now deal with the operators $\hat S_2$.
The best way to calculate the expectation value of the variational functional
\be
I[S,\tilde S]=
\bra{\ltvac} (1+\hat{\tilde S})e^{-\hat S} H e^{\hat S}
\ket{\ltvac},
\end{equation} 
is to use the usual nested commutator approach to the CCM,
\be
e^{-\hat S} H e^{\hat S} = H + [H,\hat S] + \frac{1}{2} [[H,\hat S], \hat S]+\ldots ,
\label{nested}
\end{equation}
which truncates at second order for the current Hamiltonian since the 
Hamiltonian in this form is a second order differential operator
which can act only at one or two factors in the ket state. We then evaluate
the ensuing expectation values 
by introducing complete states of intermediate states at both sides of $H$.

 We use completeness in the form
\begin{equation}
\ket{\ltvc m}=\int d\mu(\ltvc n) \ket{\ltvc n}\braket{\ltvc n}{\ltvc m},
\label{eq:complete}
\end{equation} 
where $\int d\mu(\ltvc n)$ describes the product, over the whole
lattice, of integrals over the $N$-dimensional unit sphere appropriate to the 
general $O(N)$ case. 
Relation (\ref{eq:complete}) assumes the overlap
\be
 \braket{\fuvc n_{\vec i}}{\fuvc m_{\vec i}} = \delta(\fuvc n_{\vec i}-
 \fuvc m_{\vec i}).
\end{equation}
Using the fact that the operator $\hat S_{\vec{i}\vec{j}}$, where we  now
and henceforth suppress
the pairwise subscript 2, 
projects to the right onto the ground state
for sites $\vec{i},\vec{j}$, we find that the only parts of the
commutators in Eq.~(\ref{nested}) that contribute are these where all $\hat S$ operators
are to the right of the Hamiltonian.

Using the fact, as follows from normalisation, that
\be
\braket{\fuvc n_\vec{i}}{0_\vec{i}} = \sqrt{\frac{1}{\Omega_N}},
\end{equation}
where $\Omega_N$ is the surface area of the $N$-dimensional  unit sphere,
i.e., $\Omega_4 = 2\pi^2$ for $O(4)$,
we get
\be
I[S,\tilde S] = \int d\mu(\ltvc n)\int d\mu( \{ \hat n'\})
\bra{\ltvac } 1 + \hat{\tilde S} \ket{\ltvc n}
\bra{\ltvc n} H | \{ \hat n ' \} \rangle \langle \{ \hat n ' \} |
 (1+\hat S+ \frac{1}{2} \hat S^2) \ket{0}  .
\end{equation}
If $N_g$ is the number of lattice (or grid) points, we may now write
\be\bra{\ltvc n}{\hat S}\ket{\ltvac} =
\left({\Omega_N}\right)^{-N_g/2} 
\sum_{[\vec{i}\vec{j}]} S_{\irrep{\vec{i}-\vec{j}}}
(\fuvc n_\vec{i} \cdot \fuvc n_\vec{j}),
\label{eqOpFu}
\end{equation} 
and
\be
\bra{\ltvc n}{\hat S}^2\ket{\ltvac} =
\left( {\Omega_N} \right)^{-N_g/2} 
\sideset{}{'}\sum_{[\vec{i}\vec{j}],[\vec{i}'\vec{j}']} 
S_{\irrep{\vec{i}-\vec{j}}}(\fuvc n_\vec{i} \cdot \fuvc n_\vec{j})
S_{\irrep{\vec{i}'-\vec{j}'}}(\fuvc n_{\vec{i}'} \cdot \fuvc n_{\vec{j}'})
\end{equation} 
where the prime on the sum means that only totally
disconnected diagrams are allowed, due to the projective properties
of our creation operators. The notation $\irrep{\vec{i}-\vec{j}}$ denotes
an integer irrep label; it indicates that we assume that $\hat S$ is invariant under
symmetries of the lattice, and we have to identify components at
 symmetry-equivalent lattice sites.
 Finally
\be
\bra{\ltvac}1+\hat{\tilde S}\ket{\ltvc n} =
\left({\Omega_N}\right)^{-N_g/2}
\left(1+\sum_{[\vec{i}\vec{j}]}{\tilde S}_{\irrep{\vec{i}-\vec{j}}}
(\fuvc n_\vec{i} \cdot \fuvc n_\vec{j})\right).
\end{equation} 
In a SUB2-type approximation (i.e., when the correlation operators 
$\hat {\tilde S}$ and  $\hat S$ are
 two-body operators), we can use those parts of the Hamiltonian where
only the relative variables, $\theta_{\vec{i}\vec{j}}$ defined by 
$\cos\theta_{\vec{i}\vec{j}}= 
\fuvc n_{\vec{i}} \cdot \fuvc n_{\vec{j}}$,  are present
\be
 H  \to
\sum_{[\vec{i}\vec{j}]}
\frac{-1}{\sin^2(\theta_{\vec{i}\vec{j}})} 
\partial_{\theta_{\vec{i}\vec{j}}}( \sin^2\theta_{\vec{i}\vec{j}}
    \partial_{\theta_{\vec{i}\vec{j}}})+ \sum_{\langle \vec{i}\vec{j}\rangle}
\lambda[1-\cos\theta_{\vec{i}\vec{j}}].
\label{SUB2hamil}
\end{equation}
where the kinetic operator $K$ acts twice on each relative angle 
$\theta_{\vec{i}\vec{j}}$: one from each lattice site,
$K_\vec{i}$ and $K_\vec{j}$. This explains the factor of two in the kinetic
energy terms in Eq.~(\ref{SUB2hamil}) with respect to those in the original
Hamiltonian defined in Eq.~(\ref{fullhamil}). 

\begin{figure}[tb]
\centerline{\includegraphics[height=2cm]{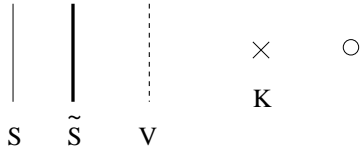}}
\caption{ The basic building blocks for the diagrammatic analysis of
the functional $I$. We denote  $S$ by a thin solid line,  $\tilde S$ by a thick
solid line and the potential by a dashed line.  A cross denotes the action of
the kinetic energy operator.  Open circles indicate that the coordinates
at these lattice points are not integrated over.}\label{fig:lines}
\end{figure}

The part of the integration measure over the 4-sphere dependent on
 $\theta$ is $\sin^2\theta\, d\theta$. Integrating over a single $S_n$ 
is the projection onto the model state, which is by construction zero.
Using this fact, we get 
the sum of all possible connected diagrams,
 generated from the basic ones sketched in Fig.~\ref{fig:lines}.
The key idea is to construct all diagrams for $I[\tilde S, S]$
 on the lattice that can be made from drawing
any of the lines ($S$, $\tilde S$, $V$) between the lattice points, subject
to the condition that the diagram is closed, no two functions $S$ can connect
at the same point, and the potential connects only nearest neighbours. Diagrams
that do not contain a potential must contain one  kinetic energy
operator $K$, and they can contain
at most one line $\tilde S$.

\begin{figure}[tb]
\centerline{\includegraphics[height=5cm]{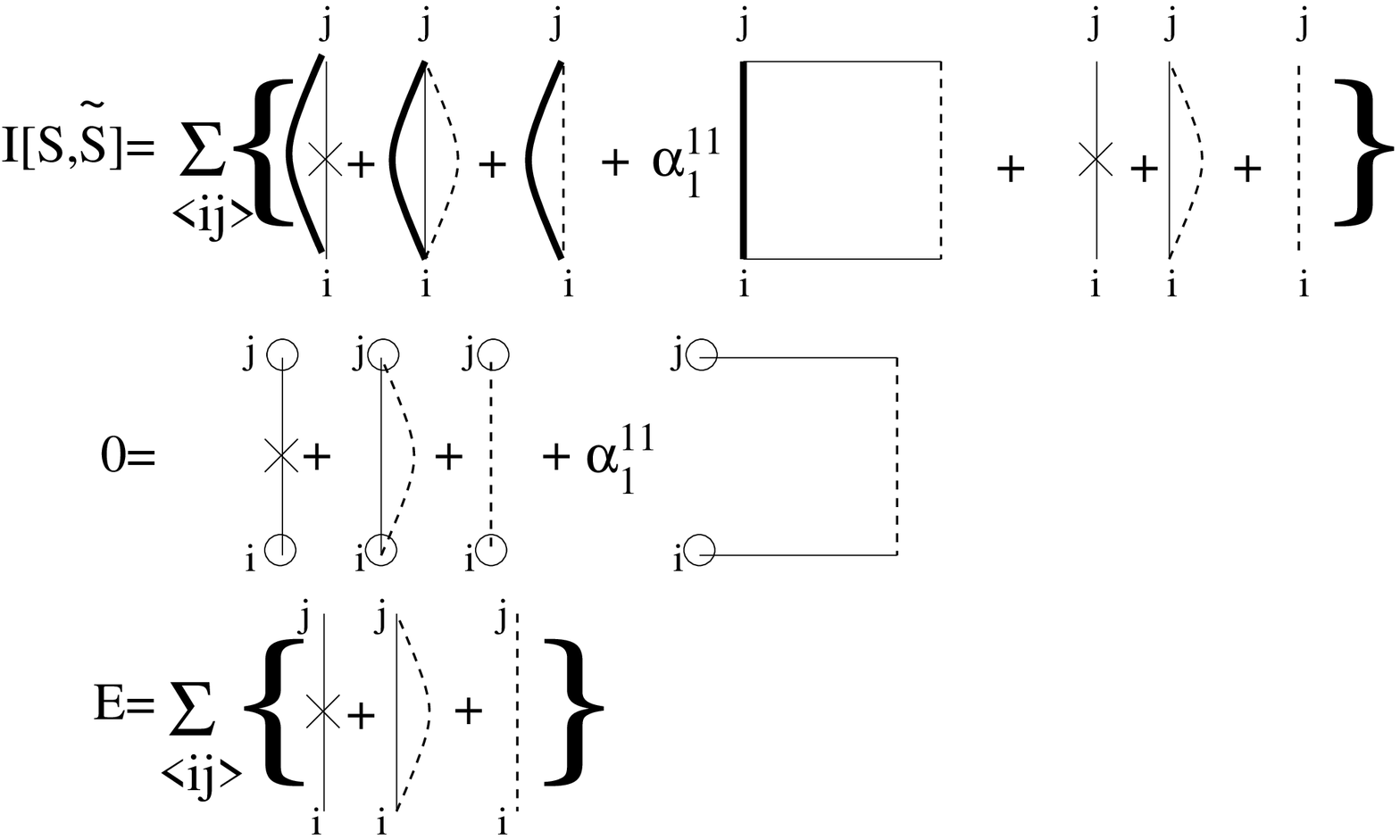}}
\caption{The diagrams for the LSUB2 approximation, together with the
non-linear equation for $S$ and the equation for the energy.  }
\label{fig:LSUB2}
\end{figure}

In the so-called LSUB2 approximation we retain only nearest-neighbour 
pairwise correlations. In Fig.~\ref{fig:LSUB2} we place the diagrams on
a lattice. Since the ket-state CCM equations are obtained through
variation of $I[S,\tilde S]$ with respect to $\tilde S_{\vec{i}\vec{j}}$, we should not
integrate over the
unit vector at these grid points. This is denoted by the
 open circles in the CCM equation. For fixed position of the
$\tilde S$ line, the square diagram in $I$ can be produced in various ways.
We choose to pick one orientation, and give it a weight $\alpha^{11}_1$
labelling the number of equivalent diagrams. The square diagram is the 
{\em only} non-local diagram in $I$, i.e., it involves nontrivial integrations
over additional lattice sites, other than $\vec{i}$ and $\vec{j}$ from
the variation with respect to $\tilde S_{\vec{i}\vec{j}}$.

Of course both the action and  the energy are infinite for an infinite
lattice. It is clear, in the LSUB2
approximation, that when the terms proportional to $\tilde S$ vanish,
which happens after minimisation, the energy can be expressed as a sum
over all nearest-neighbour links of the three diagrams shown in the last 
line of Fig.~\ref{fig:LSUB2} which do
not contain a thick line. Using the
standard assumption that each of these diagrams is independent of the 
(nearest-neighbour) sites $\vec{i}$ and $\vec{j}$ for the infinite lattice,
due to translational invariance,
 we obtain
an expression for the energy proportional to the number $N_l$ of
(nearest-neighbour) links. Therefore, in future, we shall always 
calculate the energy per
nearest-neighbour link.

A very natural basis for the expansion of all functions of $\theta$
is the set of eigenstates of the kinetic energy (i.e., to that part of 
the Laplacian depending only on the azimuthal angle). As is well 
known, these are, in any number
of dimensions, Gegenbauer polynomials or hyperspherical functions. Since
we shall use these polynomials in many places, we have dedicated 
Appendix~\ref{app:Gegenbauer} to a summary of their properties.
 For the $O(4)$ case  studied here the appropriate Gegenbauer polynomials
 $C^1_n(\cos \theta)$ are the Chebyshev polynomials of the
second kind, $U_n(\cos \theta)$.

The general equation determining $S$,  shown schematically in 
Fig.~\ref{fig:LSUB2} in the case of the LSUB2 approximation,
contains terms with zero, one, or two functions $S$ (thin lines)
linked by a potential.  The integrations over the unit-vectors at the
connecting points can be performed without any problem, including
the weight factor $1/\Omega_4=1/(2 \pi^2)$ obtained from the normalisation of the
ground state, as
\begin{eqnarray}
\lefteqn{\frac{1}{\Omega_4}\int d\mu(\fuvc n_2) S_a(\fuvc n_1 \cdot \fuvc n_2)
(\fuvc n_2 \cdot \fuvc n_3) }
\nonumber\\& =&
\frac{1}{\Omega_4}\int d\mu(\fuvc n_2) S_a(\cos \theta_{12}) 
\frac{U_1(\fuvc n_2 \cdot \fuvc n_3)}{2}
\nonumber\\& =&
\cos(\theta_{13}) \half S_{a,1}.
\end{eqnarray}
Here we have used the addition theorem for Gegenbauer polynomials,
$U_n(x) = C^1_n(x)$, as discussed in Appendix~\ref{app:Gegenbauer}. 
We have defined the generalized Fourier coefficient as
\be
S_{a,1} = \frac{2}{\pi}\int_0^\pi \sin^2\theta\, d\theta\,
2\cos(\theta) S_a(\cos \theta),
\label{genfourier}
\end{equation} 
where the factor two in front of $\cos\theta$ in the integrand of
Eq.~(\ref{genfourier}) originates from
the use of the properly normalised hyperspherical polynomial, $U_1(x)=2x$.
Since the integrand only depends on the azimuthal angle, the integration over
the other angles simply  yields a trivial constant factor,
\begin{eqnarray}
\frac{1}{\Omega_4}\int d\mu(\fuvc n_2) f(\fuvc n_1 \cdot \fuvc n_2)  & = &
\frac{1}{\Omega_4}\int  \sin^2 \theta_{12} d \theta_{12}  \sin \phi d \phi 
d \chi  f(\cos \theta_{12})  \nonumber \\
& = &
\frac{2}{\pi} \int  \sin^2 \theta_{12} d \theta_{12}  f(\cos \theta_{12})
\end{eqnarray}
as explained in more detail in Appendix~\ref{app:Quant}.

By applying the addition theorem twice we find
\be
\frac{1}{\Omega_4^{2}}\int d\mu(\fuvc n_2) \int d\mu(\fuvc n_3) \,
S_a(\fuvc n_1 \cdot \fuvc n_2) (\fuvc n_2 \cdot \fuvc n_3) S_b(\fuvc n_3 \cdot \fuvc n_4)=
\sfrac{1}{4} ( \fuvc n_4\cdot \fuvc n_1 ) S_{a,1} S_{b,1} .
\end{equation}
We thus conclude that all two- and three-line diagrams containing
one potential line are proportional to $\cos \theta$.  

\subsection{The LSUB2 approximation}

We can now
easily write down the non-linear equation for $S_1$ (i.e., the 
nearest-neighbour pairwise correlation function),
\be
Q_1 \left[
-\frac{1}{\sin^2\theta}\partial_\theta\left(\sin^2\theta
\partial_\theta S_1 \right) +\lambda (1-\cos\theta)(1+S_1)\right]
= \frac{1}{4}\lambda{\alpha^{11}_1}\cos\theta S_{1,1}^2 \ ,
\label{eq:s1_1}
\end{equation}
where $\alpha^{11}_{1}$ gives  the number of
times we can generate the box diagram for fixed position of the $\tilde S$
line, and $Q_1$ projects on all parts orthogonal to the constant function,
\be
(Q_1 f)(\cos \theta) \equiv f(\cos \theta)- \frac{2}{\pi} \int_0^\pi
 \sin^2\theta'\,d\theta' f (\ cos \theta').
\end{equation}
We rewrite Eq.~(\ref{eq:s1_1}) in terms of $\bar{S}_1=1+S_1$, to remove
the inhomogeneity from the equation, as
\be
Q_1 \left[
-\frac{1}{\sin^2\theta}\partial_\theta\left(\sin^2\theta
\partial_\theta \bar{S}_1 \right) +\lambda (1-\cos\theta)\bar{S}_1\right]
= \frac{1}{4}\lambda{\alpha^{11}_1}\cos\theta \bar{S}_{1,1}^2.
\label{eq:s1_1'}
\end{equation}
We now consistently drop the bar over $S_1$, and have therefore to remember that henceforth
 we must
satisfy the constraint
\be
\frac{2}{\pi} \int_0^\pi\sin^2\theta d\theta S_1(\cos \theta) = 1.
\end{equation} 
We can expand the solution to Eq.~(\ref{eq:s1_1'})  in terms of 
 Gegenbauer polynomials $C^1_n$, which happen to coincide
with the Chebyshev polynomials $U_n$,
\be
S_1(\cos \theta) = \sum_n S_{1,n} U_n(\cos\theta) .
\end{equation}
These polynomials satisfy the equation
\be
\frac{1}{\sin^2\theta}\partial_\theta\left(\sin^2\theta
\partial_\theta U_n(\cos\theta) \right) = -n(n+2) U_n(\cos\theta).
\end{equation} 
Using the property
\be
\int_{-1}^1 x U_n(x) U_m(x) \sqrt{1-x^2}\, dx = \half \delta_{n,m\pm1},
\end{equation} 
the equation for $S_{1}$, namely Eq.~(\ref{eq:s1_1}),
can be rewritten as the matrix equation
\be
\left(\begin{array}{llllll}
\lambda &-\half\lambda &0&0&0&\ldots\\ -\half\lambda&3+\lambda
&-\half\lambda &0&0&\ldots\\ 0&-\half\lambda&8+\lambda &-\half\lambda
&0&\ldots\\
\vdots & \vdots & \vdots & \vdots & \vdots &
\end{array}\right)
\left(\begin{array}{l} S_{1,0} \\ S_{1,1} \\ S_{1,2} \\ \vdots
\end{array}\right) =
\left(\begin{array}{l}\epsilon \\ \kappa \\0 \\ \vdots
\end{array}\right),
\end{equation} 
where $\kappa$ is a convenient shorthand for
\be
\kappa = \lambda \frac{\alpha^{11}_1}{8} S_{1,1}^2.
\label{eq:kappa}
\end{equation}
After its redefinition, we have the constraint that $ S_{1,0} =1$. 
It appears that we have one more equation than we have unknowns, until
we remember that 
the first equation (the one containing $\epsilon$) is not part of the CCM
equations; rather, it has been added by hand for elegance.
Of course we can always add an additional equation with one
additional unknown ($\epsilon$). We then choose $\epsilon$ such that
$S_{1,0}=1$, and find that we just have enough unknowns to satisfy all
equations. 

In order to solve these equations for the coefficient $S_{1,1}$ and the energy
we have to invert the matrix
\begin{equation}A = \left(\begin{array}{llllll}
\lambda &-\half\lambda &0&0&0&\ldots\\ -\half\lambda&3+\lambda
&-\half\lambda &0&0&\ldots\\ 0&-\half\lambda&8+\lambda &-\half\lambda
&0&\ldots\\
\vdots & \vdots & \vdots & \vdots & \vdots &
\end{array}\right).
\end{equation}
Obviously we need only the the top
corner coefficients of the inverse, all of which can be re-expressed in
terms of a single one of the coefficients,
\be
(A^{-1})_{12}=(A^{-1})_{21}= 2 (A^{-1})_{11} -2 \lambda^{-1}
\qquad
(A^{-1})_{22} =  4 (A^{-1})_{11} - 4 \lambda^{-1},
\end{equation}
which depend only on $\lambda$.
The top corner coefficient can be
determined through various means, e.g., by the use of
Cramer's rule:
\begin{equation}
(A^{-1})_{11} = \frac{d_{1} }{ d_0},
\end{equation}
where $d_n$ is the determinant of the matrix $A$ with the first $n$ rows
and $n$ columns removed, which satisfy the recursion relation:
\begin{equation}
d_n = [\lambda + n(n+2)]d_{n+1} - \frac{1}{4} \lambda^{2} d_{n+2}.
\end{equation}
The solutions can be expressed elegantly in the form
\begin{eqnarray}
\epsilon &=& 
\omega_{0}(\lambda)- \omega_{1}(\lambda) \kappa,\\
S_{1,1} &=& \omega_{1}(\lambda)(1 +2\kappa/\lambda).
\end{eqnarray}
Here $\omega_0$ and $\omega_1$ are elementary functions of $(A^{-1})_{11}$
\begin{eqnarray}
\omega_0(\lambda) & = & \frac{1 }{ (A^{-1})_{11}} , \\
\omega_1(\lambda) & = & - \frac{ 2 }{ \lambda (A^{-1})_{11}} + 2 =- \frac{2}{\lambda}
\omega_0+2.
\end{eqnarray}
Substituting the explicit form of $\kappa$ from Eq.~(\ref{eq:kappa}),
 we see  from the equation for the energy in
Fig.~\ref{fig:LSUB2} that at self-consistency $\epsilon=E/N_l$, where
$N_l$ is the number of nearest-neighbour links on the lattice.
We thus find
\begin{eqnarray}
E/N_l &=& \omega_{0}(\lambda)- \frac{1}{8} \lambda \omega_{1}(\lambda)
 \alpha^{11}_{1} S_{1,1}^2,
 \\ S_{1,1} &=& \omega_{1}(\lambda)(1+
\frac{1}{4}\alpha^{11}_{1} S_{1,1}^2).
\label{LSUB2quad}
\end{eqnarray} 
Equation~(\ref{LSUB2quad}) can easily be solved, and shows a perfectly regular
behaviour,
\begin{equation}S_{1,1}=\frac{2}{\alpha^{11}_1} \left(\frac{1}{\omega_1(\lambda)}
\pm \sqrt{\frac{1}{\omega_1(\lambda)^2} - \alpha^{11}_1}\right) .
\label{eq:S11}
\end{equation} 
The only acceptable solution is the one with the minus sign, since only
in that case do we have zero energy at $\lambda=0$, where $\omega_1(\lambda)
\rightarrow 0$.

\begin{figure}[tb]
\centerline{\includegraphics[width=10cm,angle=0]{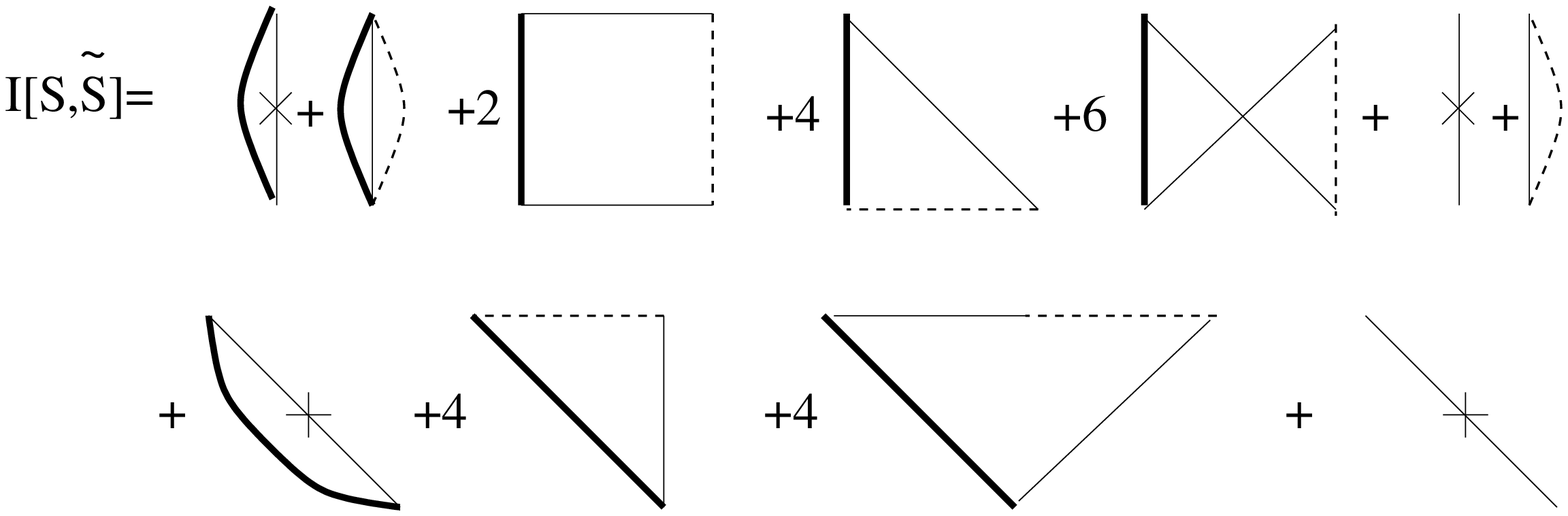}}
\caption{The diagrams for the SUB2-2 approximation in two space
dimensions. Note that because of the redefinition of $S_1 \to 1+S_1$, the two
diagrams shown in Fig.~\ref{fig:LSUB2} which contain the bare potential
line, and both a potential line and an $\tilde S_1$ line, respectively, between 
nearest-neighbour lattice points, are incorporated in the remaining diagrams.
We have also suppressed the lattice summation over the left-most link in each diagram,
which was stated explicitly in Fig.~\ref{fig:LSUB2}. }
\label{fig:SUB2-2}
\end{figure}

\subsection{The SUB2-$n$ approximations}

We now discuss what happens if we add pairwise correlations beyond nearest 
neighbour. In a SUB2-$n$ approximation we truncate at $n$ pairwise 
correlations.  Some
of the relevant diagrams are sketched in
Fig.~\ref{fig:SUB2-2} for two space dimensions in the SUB2-2 approximation
where we include only two
types of correlation functions, namely those between spins on 
nearest and next-nearest lattice
points. The additional correlations $\{S_n\ | \ {n>1} \} $, 
beyond nearest-neighbour correlations,
do not contain direct products of the correlation with the potential. 
In $S_n$, the label $n$ refers to the irrep label introduced below 
Eq.~(\ref{eqOpFu}).
Therefore they will form closed loops with the potential in which
only the lowest generalized Fourier coefficient survives, and all lead 
to differential equations of the form
\be
-\frac{1}{\sin^2\theta}\partial_\theta\left(\sin^2\theta
\partial_\theta S_n \right) = \cos \theta \ \kappa_{n}\ ; \ \ n>1,
\end{equation} 
with $\kappa_n$ a quadratic polynomial in all of the $S_{m,1}$'s.
Such an equation can easily be solved, and leads to
\begin{equation}S_{n,1}= \frac{\kappa_{n}}{6},
\label{eq:xxx}
\end{equation}
where all other Fourier-Gegenbauer 
coefficients are zero.
Equation~(\ref{eq:xxx}) is thus really a quadratic equation which only involves
the first Fourier coefficients.  

The only new diagrams contributing to
the energy are the ones involving one $S$ function and 
the kinetic energy  operator (e.g., the last diagram in Fig.~\ref{fig:SUB2-2}). Since all $\{ S_n \ | \ n>1 \}$
are proportional to the Gegenbauer polynomial of order 1, and  
are eigenfunctions of the kinetic energy,
 the derivative term is proportional
to $U_1$. When we integrate this over the relative angle we get
a zero, due to the orthogonality of the polynomials. We thus conclude that
the correlation functions $\{ S_n \  |  \ n>1 \}$
do not directly contribute to the energy;
rather they contribute only indirectly through the
additional terms due to them in the equation for $S_{1,1}$. 

In the end we obtain the coupled set of
quadratic equations
\begin{eqnarray}
\epsilon &=& \omega_{0}(\lambda) -\omega_{1}(\lambda)\lambda
\frac{\alpha^{11}_{1}}{8} S_{1,1}^2, \nonumber\\
 S_{1,1} &=&
\omega_{1}(\lambda)\left(1+ 
\frac{1}{4}\sum_{n} \alpha_{1}^{n} S_{n,1}+
\frac{1}{8}\sum_{n,m} \alpha_{1}^{nm} S_{n,1} S_{m,1} \right) ,\nonumber\\
S_{n,1} &=&
\frac{\lambda}{12}\sum_{m} \alpha_{n}^{m} S_{m,1}+
\frac{\lambda}{24}\sum_{m,l} \alpha_{n}^{ml} S_{m,1} S_{l,1}\ ;\;\;n>1.
\label{eqOpsub2}
\end{eqnarray} 
In order to understand the significance of the factors $\alpha_{l}^{nm}$ and
$\alpha_{n}^{m}$ one
should realise that we have assumed that our correlation functions are
invariant under the symmetry transformations of the lattice (which would 
appear to be a
sensible assumption for the ground state). We thus assign the same
unique numerical label to all correlation functions of such a
representation of the lattice symmetry group. The labelling function
shall be denoted by $\chi_{\vec n}$, and typically has the property that
if $| \vec n| < |\vec m|$, $\chi_{\vec n}<\chi_{\vec m}$.
The coefficients  $\alpha_{l}^{m}$ and $\alpha_{l}^{nm}$ then denote the statistical 
weights of the corresponding
diagrams, i.e., the number of times we can fit such a diagram onto the
lattice with fixed end-points. These can easily be determined from a combinatorial
calculation. We assume that the functions are invariant under the
symmetries of the cubic lattice, which leads to a great simplification.
All vectors which are related to each other by space symmetry belong
to the same irreducible representation, $\irrep{\vec{i}}$. We reserve 
$\irrep{\vec i} =1$ for the unit-length nearest-neighbour link.
The $\alpha$-coefficients may then be expressed in terms 
of their multiplicities
\begin{eqnarray}
\alpha^{n}_{i} & =&  \sum_{\vec a, \vec b, \vec c} 
 \delta_{n\irrep{\vec a- \vec b}} \delta_{i\irrep{\vec b- \vec c}}
\delta_{1\irrep{\vec c- \vec a}}, \\
\alpha^{nm}_{i} & =&  \sum_{\vec a, \vec b, \vec c, \vec d} 
 \delta_{n\irrep{\vec a- \vec b}} \delta_{i\irrep{\vec b- \vec c}}
\delta_{m\irrep{\vec c- \vec d}} \delta_{1\irrep{\vec d- \vec a}}.
\end{eqnarray}
The sums over $\vec{a},\vec{b},\vec{c},$ and $\vec{d}$ run over 
all lattice points,
subject to the restriction $\vec a \neq \vec c$, $\vec a \neq \vec d$,
and similarly for $\vec b$. These restrictions originate in the projection
part of the creation operators.

\subsection{Full SUB2 calculations}
For the case of quantum lattice spin models of relevance to
(antiferro)magnetism it has been shown that the full SUB2 equations
can often be solved by
going to Fourier space \cite{spin}. 
The corresponding equations to be solved in the present context are not very different from
those studied in Ref.~\cite{spin}.
Unfortunately, however, the $\omega_1(\lambda)$ factor in the second line
of  Eq.~(\ref{eqOpsub2}) makes it impossible to obtain usable equations 
in reciprocal space,
since the selfconsistency  condition seems to be too complicated to
allow a solution.

\subsection{Collective excitations}

An extremely important aspect of calculations of the type discussed here
is the behaviour of excitations within the model. There are many different
ways to perform such  calculations, but one of the most appealing ones is 
through the study of  small amplitude fluctuations around the ground state.
This can be formulated in several equivalent ways, but as long as we are
only interested in those excitations that have a wave function of similar
structure to the vacuum it is easy to show that this corresponds to
solving the linear random phase approximation (RPA) eigenvalue problem
\be
\int dy\,\frac{\delta^2 I}{\delta \tilde S(x) \delta S(y)} f_n(y) = E_{n} f_n(x).
\end{equation}
In the current problem this reduces to a simple matrix diagonalisation.
The matrix takes on the block structure
\be
\left(\begin{array}{cc}X_1& Y_1 \\ Y_2&X_2 \end{array}\right),
\end{equation}
with
\be
\left(X_1\right)_{ij} = \left(\begin{array}{lllll}
3+\lambda
&-\half\lambda &0&0&\ldots\\ -\half\lambda&8+\lambda &-\half\lambda
&0&\ldots\\
\vdots & \vdots & \vdots & \vdots &
\end{array}\right)_{ij} +\half \lambda \alpha^{11}_1\delta_{i,1}\delta_{j,1} S_{1,1}.
\end{equation}
In practice the matrix in this equation needs to be truncated, but
for all values of $\lambda$ truncation at order
30 seems to be more than adequate.

The matrix $Y_1$ has one non-zero row:
\be
(Y_1)_{1n} = \lambda \alpha^n_1/4 +\lambda \alpha_1^{m n} S_{m,1}/4,
\label{eq66}
\end{equation}
and $Y_2$ has one non-zero column,
\be
(Y_2)_{n1} =\lambda  \alpha_n^1/4 +\lambda \alpha_n^{1 m} S_{m,1}/4.
\end{equation}
Finally, the matrix $X_2$ has the simple structure
\be
(X_2)_{mn} = \lambda \alpha_m^n/4 +\lambda \alpha^{nl}_{m} S_{l,1}/4,
\label{eq68}
\end{equation}
where we note that in Eqs.~(\ref{eq66})--(\ref{eq68}) the summation over repeated indices is implied.
Since the RPA matrix is not symmetric, it is not obvious that its eigenvalues 
are real, and in general they can be complex.

\section{The functional form of the CCM for classical spin models}\label{sec:func}
We have already seen that
the operatorial method is driven by the potential term in the Hamiltonian,
since most of the diagrams
contain a potential line. Conversely, the functional method is 
driven by the kinetic term. The differentiations in the kinetic term 
pull correlation functions 
down from the exponential functions in Eq.~(\ref{funcfunc}) and 
link them up at the lattice point at which
the kinetic energy operator acts. The potential term simply commutes with 
the exponentials, and
acts as an inhomogeneous term in the non-linear
equations.

Since the kinetic term contains two differentiations, it links only
two correlation functions:
\begin{eqnarray}
\lefteqn{K_\vec{l} f(\fuvc n_{\vec l} \cdot \fuvc n_{\vec m}, 
\fuvc n_{\vec l} \cdot \fuvc n_{\vec k}) =} \nonumber \\
& - & D_{\vec{l}\vec{m}} f(\fuvc n_{\vec l} \cdot \fuvc n_{\vec m},
 \fuvc n_{\vec l} \cdot \fuvc n_{\vec k})  \nonumber \\
& - &
D_{\vec l \vec k} f(\fuvc n_{\vec l} \cdot \fuvc n_{\vec m},
 \fuvc n_{\vec l} \cdot \fuvc n_{\vec k})  \nonumber \\
 & - & 
\left[\fuvc n_{\vec m} \cdot \fuvc n_{\vec k} - 
(\fuvc n_{\vec l} \cdot \fuvc n_{\vec m})
( \fuvc n_{\vec l} \cdot \fuvc n_{\vec k})\right]  
f^{(1,1)}(\fuvc n_{\vec l} \cdot \fuvc n_{\vec m}, \fuvc n_{\vec l} \cdot \fuvc 
n_{\vec k}),
\end{eqnarray}
where $D_{\vec{i}\vec{j}}$ is the azimuthal part of the kinetic energy acting on the
relative angle $\theta_{\vec{i}\vec{j}}$,
\be
D f(\cos \theta ) = [-\sin^2 \theta \partial^2_{\cos \theta} + 3 \cos \theta \partial_{\cos \theta} ]f(\cos \theta ),
\end{equation}
and where $f^{(1,1)}(u,v) \equiv \partial_u \partial_v f(u,v)$.

\subsection{The LSUB2 aproximation}

In the functional method we assume, in LSUB2 approximation, that
\be
\braket{\ltvc n}{\Psi} = \exp\left(\sum_{\langle \vec{i}\vec{j} \rangle}
S(\theta_{\vec{i}\vec{j}})\right)
\end{equation}
We find that the CCM variational functional can now be written in
the LSUB2 scheme as
\begin{eqnarray} 
I(S, \tilde S)= N_l
\frac{2}{\pi} \int \sin^2\theta d\theta
\left[(1+\tilde S(\theta))\right]
\left[ -\left(S''+2 \cot(\theta)S'+(S')^2\right)
+V(\theta)\right],
\end{eqnarray} 
which only involves one-link quantities. We get no terms equivalent to the square diagram in the functional method,
since the {\em function} $S$ commutes with the potential.
As before, all open
diagrams disappear, even the one where the open line contains a derivative.
For example, one of those diagrams consists of two functions $S'$ linked by an $\tilde S$, which form
a closed triangle formed by linking two
correlation functions, from the kinetic term acting at lattice point $\vec i$,
with  $\tilde S_{\langle \vec{i}, \vec{k}\rangle}$.
It has a value proportional to 
\begin{equation}
J \equiv
\int {\rm d}^4 \fuvc n_{\vec{i}}  {\rm d}^4 \fuvc n_{\vec{k}} {\rm d}^4 \fuvc n_{\vec{l}}
S'(\fuvc n_{\vec{i}} \cdot \fuvc n_{\vec{k}})
S'(\fuvc n_{\vec{i}} \cdot \fuvc n_{\vec{l}})
[(\fuvc n_{\vec{i}} \cdot \fuvc n_{\vec{k}})
(\fuvc n_{\vec{i}} \cdot \fuvc n_{\vec{l}})-
(\fuvc n_{\vec{k}} \cdot \fuvc n_{\vec{l}})]
\tilde S(\fuvc n_{\vec{i}} \cdot \fuvc n_{\vec{k}}) .
\label{intJ}
\end{equation}
We now choose a coordinate system where the axes are oriented such that
\begin{eqnarray}
\fuvc n_{\vec{i}} = \begin{pmatrix} 1 \cr 0 \cr 0 \cr 0 \end{pmatrix},\quad
\fuvc n_{\vec{k}} = \begin{pmatrix} \cos \theta_2  \cr \sin \theta_2 \cr 0 \cr 0\end{pmatrix}, \quad
\fuvc n_{\vec{l}} = \begin{pmatrix} \cos \theta_3 \cr \sin \theta_3 \cos \phi_3 \cr
 \sin \theta_ 3 \sin \phi_3 \cr 0\end{pmatrix}.
\end{eqnarray}
The integration  in Eq.~(\ref{intJ}) then factorises into the product of  two 
integrations over the relative angles,
 with one integration over the $\fuvc n_\vec{i}$,
\begin{eqnarray}
J & = &  - \int d\mu (\hat n_{\vec{i}})  
\int \prod_{j=2}^3 \sin^2 \theta_j \sin \phi_j {\rm d}
 \theta_j {\rm d} \phi_j {\rm d} \chi_j \nonumber \\ \qquad
& & 
\times S'(\cos \theta_2) S'(\cos \theta_3) (\sin \theta_2 \sin \theta_3 \cos \phi_3)
 \tilde S( \cos \theta_2) = 0  ,
\end{eqnarray}
which vanishes due to the trivial zero that arises from the integration
over $\phi_3$.

For the special case of the LSUB2 approximation
it is better to rewrite the expression explicitly
in terms of the exponentiated quantities,
\begin{eqnarray}
I(S, \tilde S)&=& N_l
\frac{2}{\pi} \int \sin^2\theta d\theta
\left[\{ 1+\tilde S(\theta) \} \exp \{ -S(\theta)\} \right]\nonumber\\&&
\times
\left[- \frac{1}{\sin^2\theta}\frac{d}{d\theta}\sin^{2}\theta \frac{d}{d\theta}\exp\{S(\theta)\}+
V(\theta)
\exp\{ S(\theta) \} \right].
\end{eqnarray} 
After redefinition of $S$ and $\tilde S$,
\begin{equation}u=\exp(S),\;\; v= (1+\tilde S)\exp(-S)\end{equation} 
we have a standard variational functional
\begin{eqnarray} I(u,v)&=& N_l
\frac{2}{\pi} \int \sin^2\theta d\theta\,
 v(\cos\theta)\nonumber\\&&
\left[-
\frac{1}{\sin^2\theta}\frac{d}{d\theta}\sin^{2}\theta \frac{d}{d\theta}+
\lambda(1-\cos\theta)-\frac{E}{N_l}\right]
u(\cos\theta).
\end{eqnarray} 
in which we have now introduced the energy $E$ as a Lagrange multiplier since we have lost the
intrinsic normalisation from the exponentials.
The equations for $u$ and $v$, which are obtained by variation with respect 
to $v$ or $u$, respectively, are now identical (since the operator 
is self-adjoint)
and we have to solve the equation
\be
\left[-
\frac{1}{\sin^2\theta}\frac{d}{d\theta}\sin^{2}\theta 
\frac{d}{d\theta}+
\lambda(1-\cos\theta)-\frac{E}{N_l}\right]
u(\cos\theta)=0.
\end{equation}
We substitute $u(\cos\theta)= y(\theta/2)/\sin(\theta)$,
 $\theta/2=\psi$ and obtain 
Mathieu's equation
\be
y'' + \left(4\left\{\frac{E}{N_l} - \lambda + 1\right\} 
-2\left[-2 \lambda \right]\cos 2\psi \right) y = 0.
\end{equation}
The only problem that remains is what boundary conditions we need to impose.
Clearly the original function $u$ depends on $\cos \theta $, and
must be even in $\theta$. Since we had to 
divide the function $y$ by $\sin\theta$ to get $u$, we find that $y$ must
be odd in $\theta$.
This immediately fixes
the lowest eigenstate to occur at
\be
\frac{E}{N_l} = \lambda - 1 + \frac{1 }{ 4}b_2(-2\lambda) , 
\end{equation}
where
$b_2$ is the lowest odd characteristic value of Mathieu's equation with period $\pi$ \cite{AS64}. Notice that
we also get an estimate of the excitation energies for free,
by replacing $b_{2}$ by $b_{2n}$.
 This reflects the fact that in
this case the result is fully variational, and could equally well have 
been obtained by performing a variational calculation,
\be
\delta E =0 \ ; \ \ E = \frac{\bra{\psi}H\ket{\psi}}{\braket{\psi}{\psi}},
\end{equation}
with the sum-over-links Ansatz for the wave function,
\be
\braket{\ltvc n} {\psi} = \sum_{\langle \vec i \vec j\rangle } e^{S_{\vec i \vec j}}.
\label{varsum}
\end{equation}
We note that the corresponding Jastrow-type variational calculation of Chin 
\cite{Shi97} differs from Eq.~(\ref{varsum}) only by the use of a product
of exponentials instead of the sum of exponentials used here.
This relation  of the functional form of the CCM 
with a variational calculation is an accident of the LSUB2 
approximation; it does not persist to other approximations in the SUB2-$n$
or higher schemes.

\subsection{The SUB2-$n$-$m$ approximations}
As in the case of the operatorial form we also wish to incorporate 
additional longer-range correlations
of two point nature. In this case we need to truncate both on the number $n$
of such longer-range functions as above, but also on the number $m$ of basis functions
in which each such SUB2-$n$ correlation function is expanded. The resulting approximation
scheme is called the SUB2-$n$-$m$ scheme.

For the general SUB2-$n$-$m$ case, where we assume functional dependence on
an irreducible set of correlation functions, we cannot use the simplifying 
technique
sketched above, but we shall have to deal with the full complexity of
the problem. Since the reference
state is an eigenstate of the sum of all the kinetic energy operators, thus
corresponding to  the weak-coupling  limit, a natural set of functions is the
set of Gegenbauer polynomials, as will be seen below.

We expand $S$ as
\begin{equation} 
S = \sum_{[{\vec i}{\vec m}]}  S_\irrep{\vec m}(\fuvc n_{\vec i}
\cdot \fuvc n_{\vec i+\vec m}),
\end{equation} 
where both sums run over all lattice sites. We will
introduce the specific details and the exact specification
 of our basis functions later. For $\tilde S$ we
use the Gegenbauer polynomials we introduced earlier,
\begin{equation}
\tilde S = \sum_{[\vec i\vec m]}
 \sum_{k=1}^\infty \tilde s_{\vec m,k} U_k(\fuvc n_{\vec i}
\cdot \fuvc n_{\vec i+\vec m}).
\end{equation} 
If we insert these correlation functions, $S$ and $\tilde S$, into
our functional $I$ we find:
\begin{eqnarray}
\frac{\delta I }{ \delta \tilde s_{\vec m,k}} &=& 
\langle 0 |
U_k( \fuvc n_0 \cdot \fuvc n_{\vec m}) \biggl [
-\frac{1}{\sin^2\theta}\partial_\theta\left(\sin^2\theta
\partial_\theta S_{\irrep{\vec m}} \right)
\nonumber\\&&
- (1 - ( \fuvc n_0
\cdot \fuvc n_{\vec m})^2) S'_\irrep{{\vec m}} S'_\irrep{\vec m} 
\nonumber\\&&-
\sum_{\vec m'}
(\fuvc n_0 \cdot \fuvc n_{\vec m} - (\fuvc n_{\vec m'} \cdot \fuvc n_{\vec
m})(\fuvc n_0 \cdot \fuvc n_{\vec m'}) S'_{\irrep{\vec m- \vec m'}} S'_{\irrep{\vec m'}}
\nonumber\\&& +
\delta_{1\irrep{ \vec m}} V_{\vec m} \biggr ] | 0 \rangle,
\end{eqnarray}
The potential function $V_{\vec{i}-\vec{j}}$ between nearest-neighbour
sites  can be expressed in Gegenbauer polynomials as
\begin{equation} 
V_{\vec{i}-\vec{j} } =  \lambda U_0(\fuvc n_\vec{i} \cdot \fuvc n_{\vec j} )
- \frac{1 }{ 2} \lambda U_1(\fuvc n_\vec{i} \cdot \fuvc n_{\vec j} ).
\end{equation} 
Because the equations are translationally invariant we restrict ourselves
to correlations between the origin, $\vec{i}=0$, and the other lattice sites.
Since only the
derivative of $S$ appears in the expression it is more convenient to
parametrise this derivative directly in terms of Gegenbauer polynomials, 
rather than by parametrising $S$ itself. This may readily be archieved,
 since a truncation on $S$ in terms of Gegenbauer
polynomials implies a truncation of $S'$ one order lower. Thus, we write
\begin{equation}
S'_{\irrep{\vec m}} = \sum_{k=0}^\infty s_{\irrep{\vec m},k}
U_k(\fuvc n_0 \cdot \fuvc n_{\vec m}).
\end{equation} 
In this case  the sum over the Gegenbauer polynomials
starts at the constant function, $U_0$, since $S'$ unlike $S$ can
contain a constant part.

A tedious but straightforward calculation gives the following
expressions for the variational equations, 
\begin{eqnarray} 
0 & = & 
\int \frac{d\mu(x) }{ \Omega_4} U_k (x) \Big[ \sum_{j=1}^\infty \frac{1
}{ 2} ( -j s_{\irrep{\vec n},j+1} + (j+2) s_{\irrep{\vec n},j-1}) U_j (x) \nonumber\\
& - &
\frac{1 }{ 2} \sum_{j=0}^\infty \sum_{l=1}^\infty s_{\irrep{\vec n},l} (s_{\irrep{\vec
n}, j+l} - s_{\irrep{\vec n},j+l+2}) U_j(x) - \frac{1 }{ 4} \sum_{j=0}^\infty
\sum_{i=0}^j s_{\irrep{\vec n}, j-i} s_{\irrep{\vec n}, i} U_j(x)\nonumber\\
& +& \frac{1 }{ 4}
\sum_{j=0}^\infty \sum_{i=0}^{j-2} s_{\irrep{\vec n}, j-2-i} s_{\irrep{\vec n},i}
U_j(x) \nonumber\\
& - & \frac{1 }{ 4}
\sum_{\vec m} \sum_{j=0}^\infty \left( \frac{(j+2) s_{\irrep{\vec n-\vec m},j-1}
 s_{\irrep{\vec m},j-1} }{ j(j+1)} +
\right. \nonumber\\ 
&&\left.\qquad \frac{j s_{\irrep{\vec n - \vec m},j+1} s_{\irrep{\vec
 m},j+1} }{ (j+1)(j+2)} - \frac{s_{\irrep{\vec n - \vec m},j-1} s_{\irrep{\vec
 m},j+1}}{ (j+1)^2} - \frac{s_{\irrep{\vec n - \vec m},j+1} s_{\irrep{\vec m},j-1}}{ (j+1)^2}
 \right) U_j(x) \nonumber\\
& + & \lambda (  U_0(x) - \frac{1 }{ 2}U_1(x)) \Big]  ,
\end{eqnarray} 
where $x= \fuvc n_0 \cdot \fuvc n_{\vec n}$, and where we have used the
addition theorem for Gegenbauer polynomials.  In order to solve these equations
 we  also use the orthogonality relation for the Gegenbauer polynomials:
\begin{equation}
\int \frac{d\mu(x) }{ \Omega_4} U_i(x) U_j(x)  = \delta_{ij}.
\end{equation}
Note that the integration measure contains the normalisation constant from
integration over the other (than the azimuthal) angles of the four-dimensional 
unit sphere.
Of course in practical calculations
the truncations of the infinite sums are important. These are performed by
assuming that the coefficients $s_{\irrep{\vec n},j}$ vanish for $j$ larger 
than a cut-off value $m$.

Once we have done that we can determine the energy per
nearest-neighbour link from the expression, 
\begin{eqnarray}
E/N_l & = & 
1/(2D) \sum_{\vec n} \Big[ 
- \frac{1 }{ 2} \sum_{l=1}^\infty s_{\irrep{\vec n},l} (s_{\irrep{\vec
n}, l} - s_{\irrep{\vec n},l+2})  - \frac{1 }{ 4}  
 (s_{\irrep{\vec n}, 0})^2  
+  \lambda  \Big] ,
\end{eqnarray} 
where $D$ represents the number of spatial dimensions.

\subsection{Collective excitations}

One way to calculate excitations is to use the so-called Feynman technique.
Thus, the excitation energy for the state $\ket{\Psi_e} \equiv X \ket{\Psi}$ can be evaluated by
the double commutator of the Hamiltonian with the excitation operator,
if we assume that  $X \ket{\Psi}$ is an eigenstate of the Hamiltonian.
In the CCM one often uses the same approach, even though it is no longer
true that the bra and ket excited states are generated by the same
excitation operator, $X$, and thus the approximations in this scheme are not
easily controlled.  For the case of the functional form of the CCM
we use the operator that generates the first excited state in the limit
of zero coupling constant. 
We can calculate the energy gap in 
this approximation analytically, knowing only the properties of the ground
state.

As usual in the functional form, the truncation is performed directly on 
the wave function rather than on the operators, and we assume  the SUB2 form
for the ground-state correlation functions and the LSUB2 form for the excitation
function,
\begin{eqnarray}
\braket{\ltvc n}{\Psi_e} &=& X \exp\left(\sum_{[\vec i\vec j]}S(\cos \theta_{\vec i\vec j})\right),
\nonumber\\
\braket{\smash[t]{\tilde \Psi_e}}{\ltvc n} &=&
 \left(1+\sum_{[\vec{i}\vec{j}]}\tilde S(\cos \theta_{\vec i\vec j})\right)
\exp\left(-\sum_{[\vec i\vec j]}S(\cos \theta_{\vec i\vec j})\right) X,
\end{eqnarray}
where the excitation function takes the LSUB2 form,
\begin{equation} 
X = \sum_{\langle \vec i \vec  j\rangle} \fuvc n_{\vec i} \cdot \fuvc n_{\vec j}.
\end{equation} 
We now  evaluate the expectation value of the second commutator of
the excitation function $X$ with the Hamiltonian, and find the usual result
for the excitation energy,
\be
\Delta E =-\frac{1 }{ 2} \left( \int \frac{d \mu (\{ \fuvc n \}) }{ \Omega_4^{N_g}}
 (1+\tilde S)e^{-S} [[H,X],X] e^S  \right)
\left( \int \frac{d \mu (\{ \fuvc n \}) }{ \Omega_4^{N_g}} (1+\tilde S) X^2 \right)^{-1}
\end{equation}
The numerator is the second derivative of the functional $I[\tilde
S,S]$ with respect to $s_{1,1}$. The denominator can be represented by
a set of diagrams of one link terms, with $\cos^2 \theta_{\vec{i}\vec{j}}$ and
triangles containing two one-link terms, $\cos \theta_{\vec{i}\vec{k}} \cos
\theta_{\vec{k}\vec{j}}$, and $\tilde S$.

We can also calculate the excitations from a small-fluctuation RPA-type
 expansion
around the ground state, similar to the method used in the operatorial
approach. The best way to derive such an approach is by
studying the small-amplitude limit of the classical equations of motions
that can be derived from the quantum action functional
\begin{eqnarray}
{\cal A}[\tilde S, S] & = &\int_0^T d t \int \frac{d\mu( \{ \fuvc n \}) }{ \Omega_4^{N_g}} 
(1+ \tilde S){\rm e}^{-S}[i \partial_t - H ] {\rm e}^S \nonumber \\
& = & -i \int_0^T d t \int \frac{d\mu( \{ \fuvc n \}) }{ \Omega_4^{N_g}}
  \dot {\tilde S} S - \int_0^T d t\, I[\tilde S, S].
\end{eqnarray}
If we expand $\tilde S$ and $S$ in Gegenbauer polynomials, we obtain an
action of standard canonical form,
\begin{eqnarray}
{\cal A}[\tilde S, S] 
& = & -i \int_0^T d t 
  \sum_{\vec n, i} {\dot {\tilde s}}_{{\vec n},i} s_{{\vec n},i} - \int_0^T d t\, I[\tilde s, s].
\end{eqnarray}
We thus need to change 
the expansion of $S'$ in polynomials, to an expansion of $S$. 
This is a straightforward calculation
 using the recursion relation of the polynomials. In order
to establish orthogonality we had to use the fact that $S$ had SUB2 form,
 so that we could change to relative variables. For more complicated 
correlations this is no longer possible, and we lose the simple form of the
time-dependent variational principle. 

By linearising the Hamiltonian equation of motion derived from ${\cal A}$ we
see that we only need 
consider the symmetric second-order expansion of $I$ around its 
stationary ground-state value,
\begin{eqnarray}
I[\tilde S, S] & \approx & I[\tilde S_0, S_0] + \frac{\delta^2 I[\tilde S, S] }{ 
\delta \tilde S \delta S}\Big|_{\tilde S_0,S_0} \delta \tilde S \delta S 
\nonumber \\
& \approx & I[\tilde S_0, S_0] + \frac{\delta^2 I[\tilde S, S] }{
\delta \tilde S \delta S'}\Big|_{\tilde S_0,S'_0} \delta \tilde S (T \delta S),
\label{funcRPA}
\end{eqnarray}
where we have introduced the matrix $T$ (and see Eq.~(\ref{Gegder}) in 
Appendix~\ref{app:Gegenbauer}) which implements the change 
in basis when
we expand $S'$ or $S$ in terms of Gegenbauer polynomials.
 The lowest eigenvalue of 
the functional varied with respect to $\tilde S$ and $S$ is the lowest 
excitation energy.

\section{Solution and Results}\label{sec:SolRes}

\subsection{Numerical methods} 

Apart from standard matrix diagonalisations and inversions, the most
challenging numerical problem encountered in the present work is
tracing a solution to a parametric set of non-linear equations, where the
coupling constant $\lambda$ is the running parameter. This problem
has been well studied by numerical analysts, since it occurs in
areas of economics and engineering as well as in physics. There
are various ways to solve such a problem, but one of the most robust
and stable ones is based on a technique developed by
Rheinboldt and collaborators \cite{Rheinboldt}. This technique
combines a predictor-corrector method with a Newtonian method to solve
systems of non-linear equations, as one steps through parameter space
following the solution.  We have used the implementation of this scheme
in the PITCON
program developed by Rheinboldt and Burkardt \cite{Pitcon}.

\subsection{LSUB2 Results}
\begin{figure}[tb]
\centerline{\includegraphics[height=7cm]{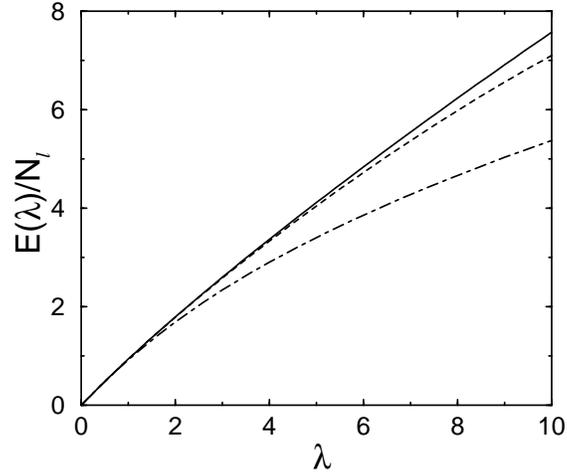}}
\caption{The energy per link in LSUB2 approximation, obtained in 2D (solid line)
and 3D
(dashed line) from the CCM operatorial form, as well as by the solution to the Mathieu
equation (dashed-dotted line) from the CCM functional form.}
\label{fig:E_LSUB2}
\end{figure}

For this approximation there is a large difference between the functional and operatorial
methods. Even though the functional method has the same solution
for any spatial dimensionality, and is explicitly described by the solution to 
the Mathieu equation,
this is not true for the operatorial method, where the result depends on
the spatial dimensionality through the statistical weight of the square
diagram. Substituting 
$\alpha^{11}_{1}=2$ for two dimensions, and $\alpha^{11}_1=4$ for three
dimensions, we find the results for the ground-state energy sketched 
in Figure~\ref{fig:E_LSUB2}.  Both of these operatorial CCM solutions actually 
terminate and return. 
The termination points  which occur for larger values than shown in Fig.~\ref{fig:E_LSUB2} are
where $\omega_1 = 1/\sqrt{\alpha^{11}_1}$. In one spatial dimension the 
non-local diagram does not exist, and hence there is no termination point 
in this case. 
However, in two and three spatial dimensions the termination point is 
at $\lambda = 115.2$ and $\lambda = 16.3$ respectively.
By contrast, the functional
form does not terminate and turn around. Rather, the solution exists for all $\lambda$
in this case, and hence no termination point exists.

\subsection{SUB2-$n$(-$m$) Results}

For the operatorial form we need to solve the non-linear Equations~(\ref{eqOpsub2}). Using
the PITCON approach mentioned above, 
the equations can easily be solved once we determine the functions 
$\omega_0(\lambda)$ and $\omega_1(\lambda)$,
by truncating the set of equations where we keep only the lowest 
$n$ coefficients $S_{k,1}$ with $k=1,2,\cdots, n$ 
(and set the higher terms with $k>n$ to zero).
 This approach 
is usually called the SUB2-$n$ approximation. There is no real 
computational restriction
on the value of $n$ used in this approach, apart from the
generation of the $\alpha$-coefficients, which becomes very time-consuming for 
large values of the truncation index $n$. As we shall see below, convergence is attained 
before that becomes a real burden.

In the functional method we need to truncate on the number of different
$S$ functions, corresponding to the set of distinct lattice vectors 
retained (i.e., those which differ under lattice symmetries), and for 
each distinct
pairwise correlation function thereby retained we also
have to truncate further on the number of basis functions in which they are expanded. 
Generally, the corresponding SUB2-$n$-$m$ results depend more on the the number 
$n$ of pairwise correlation
components taken into account than on the number $m$ of basis functions 
are used for the expansion of each component. Even close to the termination point all 
components behave smoothly and are not very large.  They are 
well approximated with just three basis functions.

\begin{figure}[htb]
\centerline{\includegraphics[height=7cm]{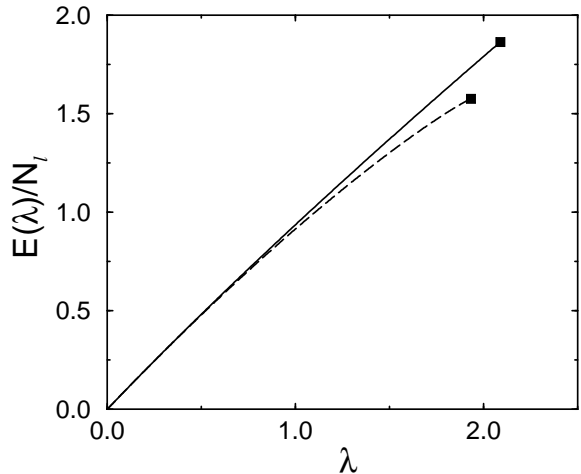}}
\caption{The energy per link in the SUB2-$n$ (operatorial, solid line) and 
SUB2-$n$-$m$ (functional, dashed line) approximations in 1 space dimension.
Both schemes exhibit a termination point, which is denoted by the solid 
squares. The results are shown for sufficiently high values of $n$ and $m$ for convergence to have been attained on the scale of the figure.}
\label{fig:Etot_1d}
\end{figure}
\begin{figure}[htb]
\centerline{\includegraphics[height=7cm]{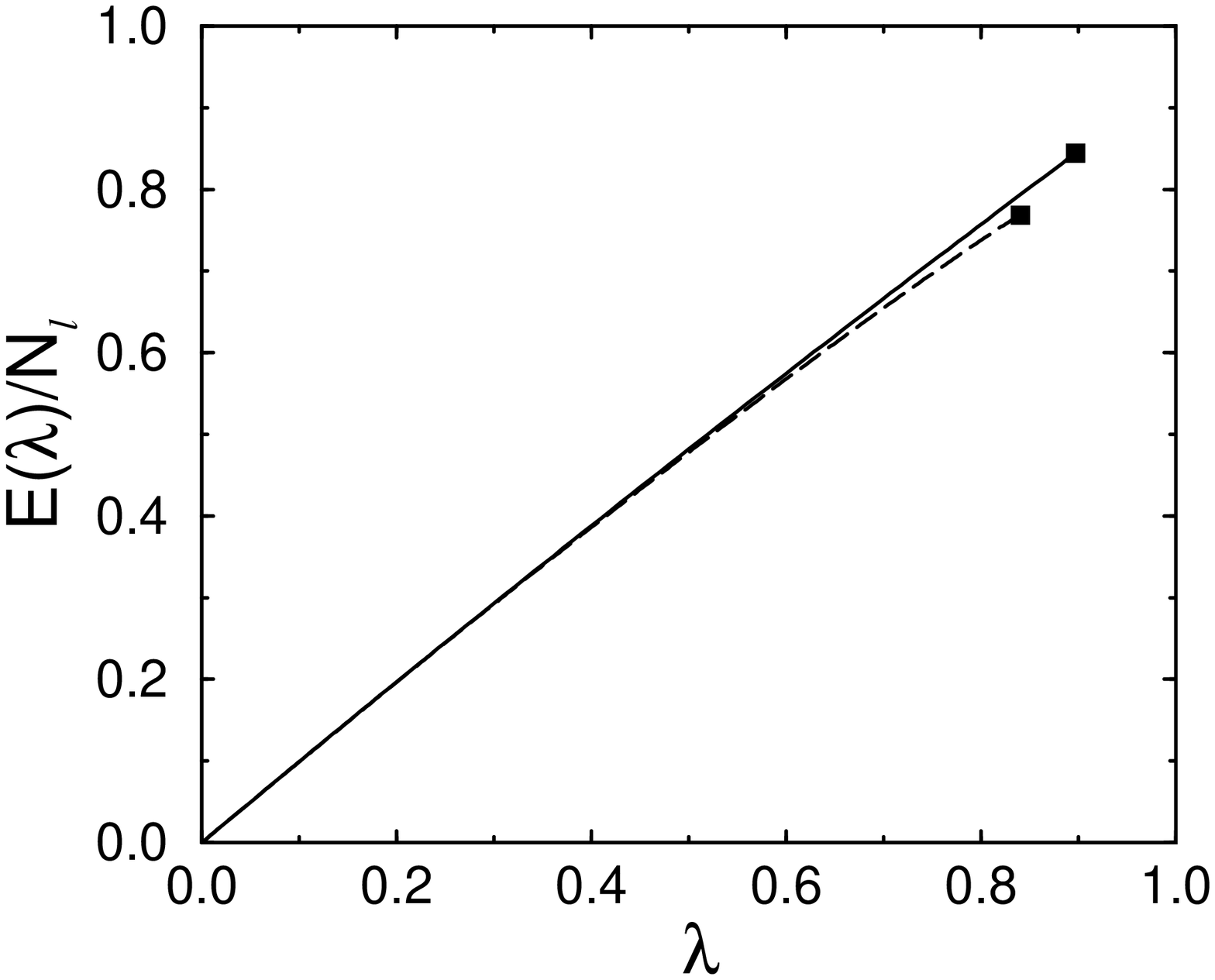}}
\caption{The energy per link in the SUB2-$n$ (operatorial, solid line) and 
SUB2-$n$-$m$ (functional, dashed line) approximations in 2 space dimensions.
Both schemes exhibit a termination point, which is denoted by the solid squares.
The results are shown for sufficiently high values of $n$ and $m$ for convergence to have been attained on the scale of the figure.}
\label{fig:Etot_2d}
\end{figure}
\begin{figure}[htb]
\centerline{\includegraphics[height=7cm]{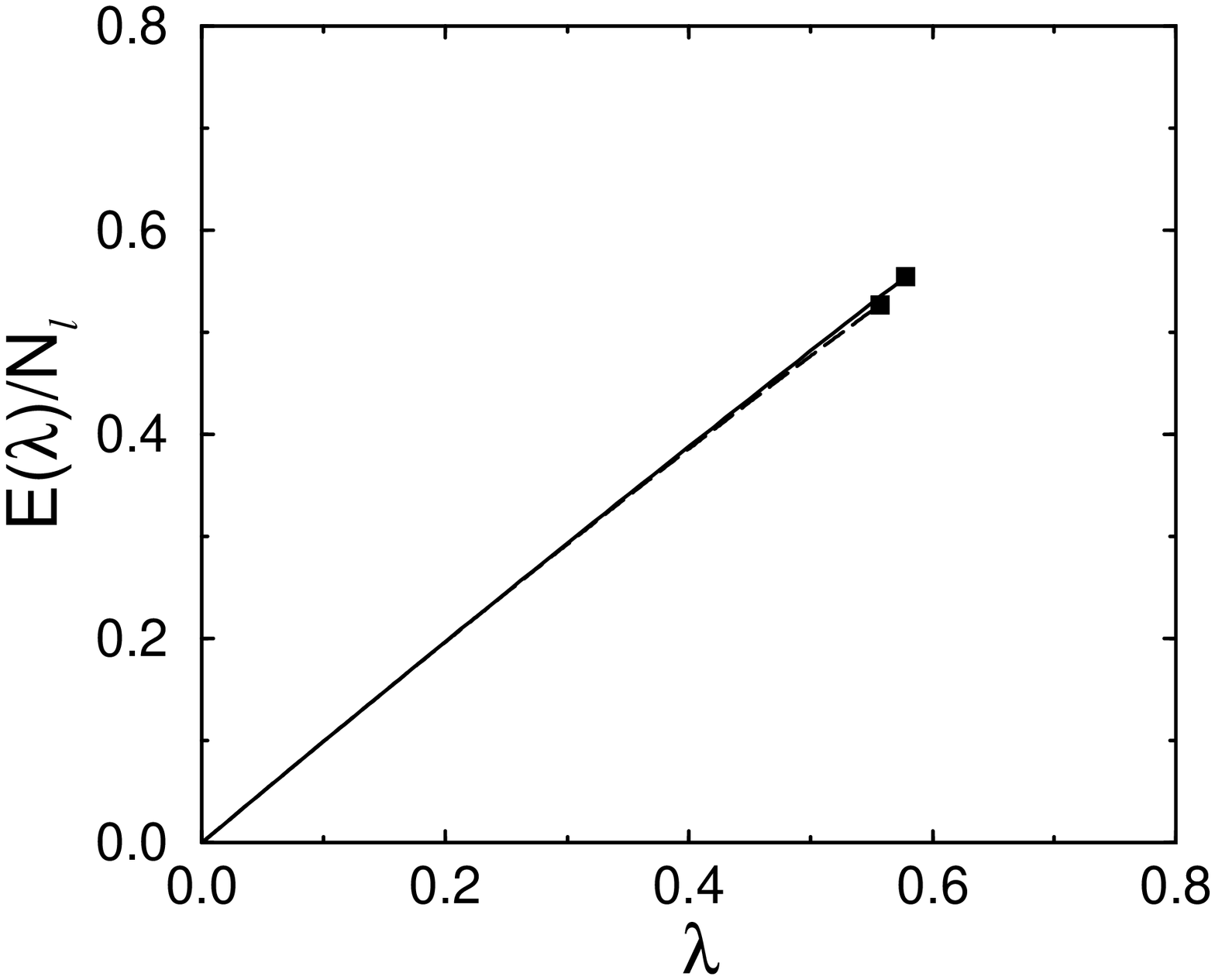}}
\caption{The energy per link in the SUB2-$n$ (operatorial, solid line) and 
SUB2-$n$-$m$ (functional, dashed line) approximations in 3 space dimensions.
Both schemes exhibit a termination point, which is denoted by the solid squares. The results are shown for sufficiently high values of $n$ and $m$ for convergence to have been attained on the scale of the figure.}
\label{fig:Etot_3d}
\end{figure}

We have studied the behaviour of the energy in 1, 2 and 3 space dimensions,
and we present some relevant results in 
Figs.~\ref{fig:Etot_1d}--\ref{fig:Etot_3d}. In each of these figures we 
compare the functional to the operatorial approach, both truncated at
such high values of $n$ and $m$ that the results are converged on the scale
shown in the diagrams. 
We find, following the path of the solution, that for all three cases the
solutions are not single-valued in $\lambda$. Each solution actually consist of two
branches, that lie closer and closer together as we increase the order $n$.
One of the two branches is ``unphysical'' in the sense that it
does not connect directly with with the known solution for small coupling
constants. This is also borne out by the excitation energies, as we see below.
The mathematical reason for this behaviour is obvious. Thus, roots of non-linear
equations cannot disappear; they have to collide with another root. Physically
we interpret this behaviour as a sign of a break-down point of our
approximation, and a possible indication of a phase transition at the
corresponding termination point. The strength
of this latter assumption is of course weakened by the fact that we find
a phase transition here for the one-dimensional model where no such physical phase 
transitions occur \cite{MW66}. 
The fact that our putative phase transition occurs at smaller 
and smaller values of $\lambda$ as we increase the number of spatial 
dimensions may  indicate
that the occurrence of such a transition is more plausible in higher 
dimensions.

\begin{figure}[htb]
\centerline{\includegraphics[height=7cm]{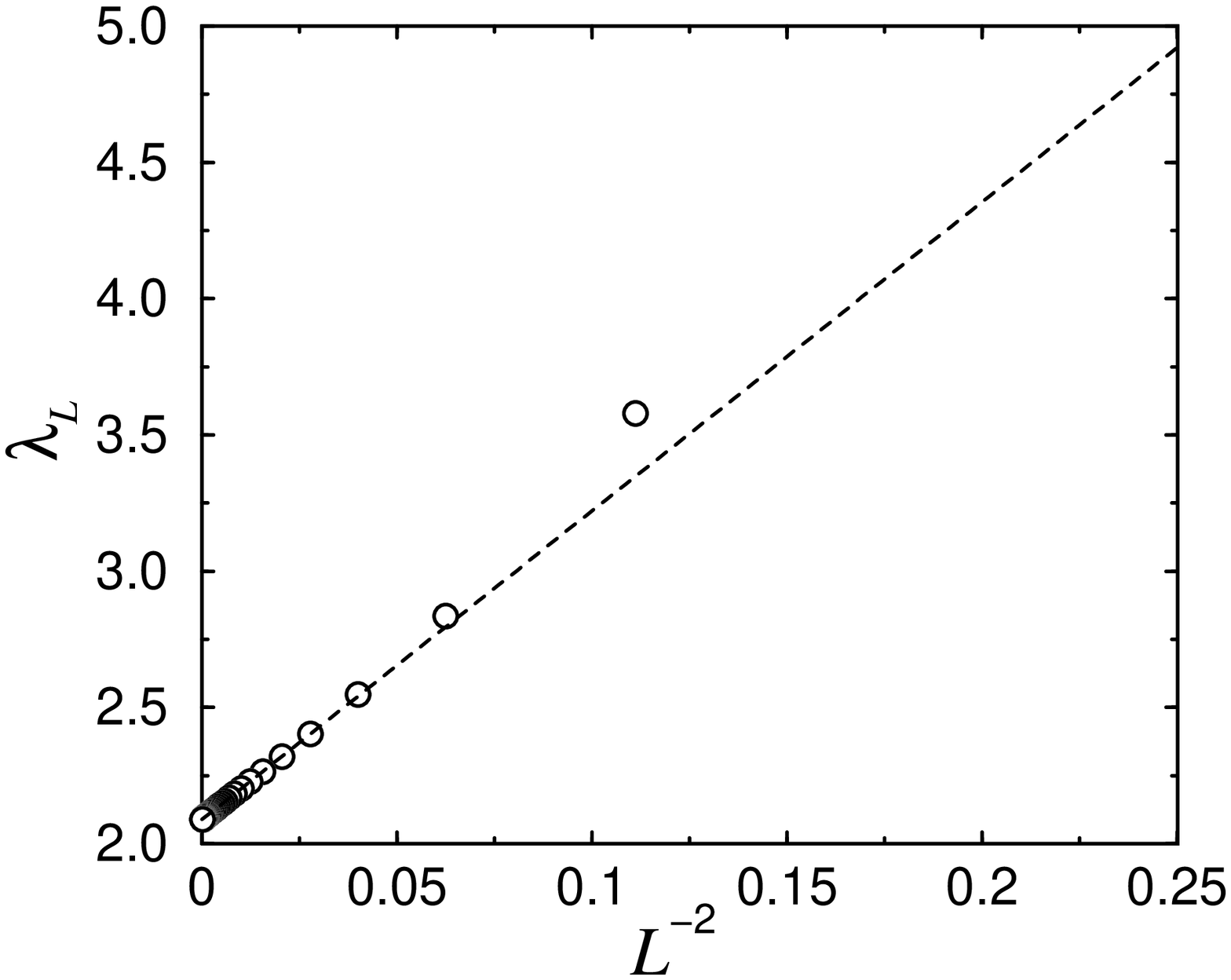}}
\caption{The SUB2 termination point as a function of $L^{-2}$ defined in Eq.~(\ref{lambda}) 
in 1 space dimension, using the CCM operatorial form.}
\label{fig:lam_end_1d}
\end{figure}

\begin{figure}[htb]
\centerline{\includegraphics[height=7cm]{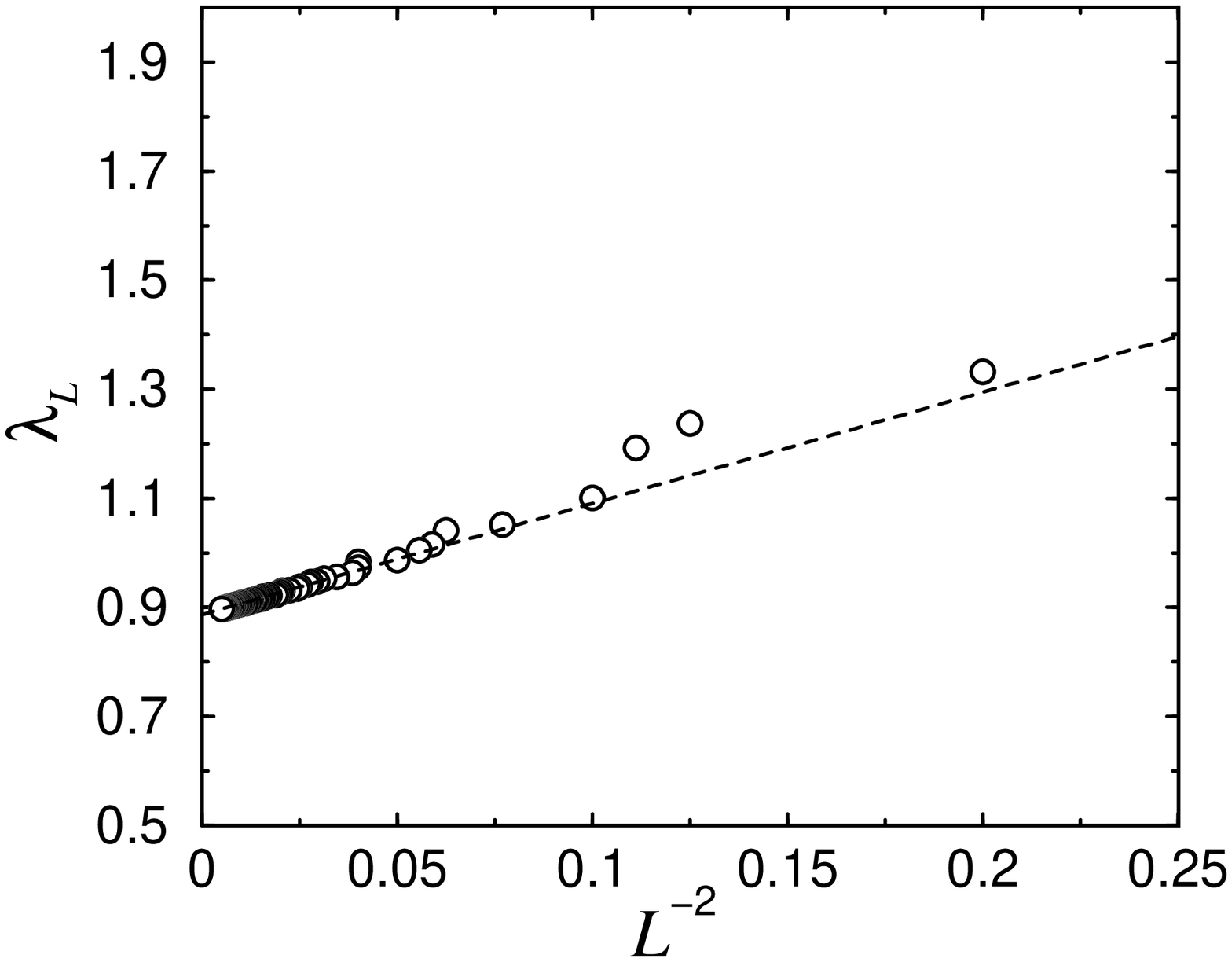}}
\caption{The SUB2 termination point as a function of  $L^{-2}$ defined in Eq.~(\ref{lambda})
in 2 space dimensions, using the CCM operatorial form.}
\label{fig:lam_end_2d}
\end{figure}

\begin{figure}[htb]
\centerline{\includegraphics[height=7cm]{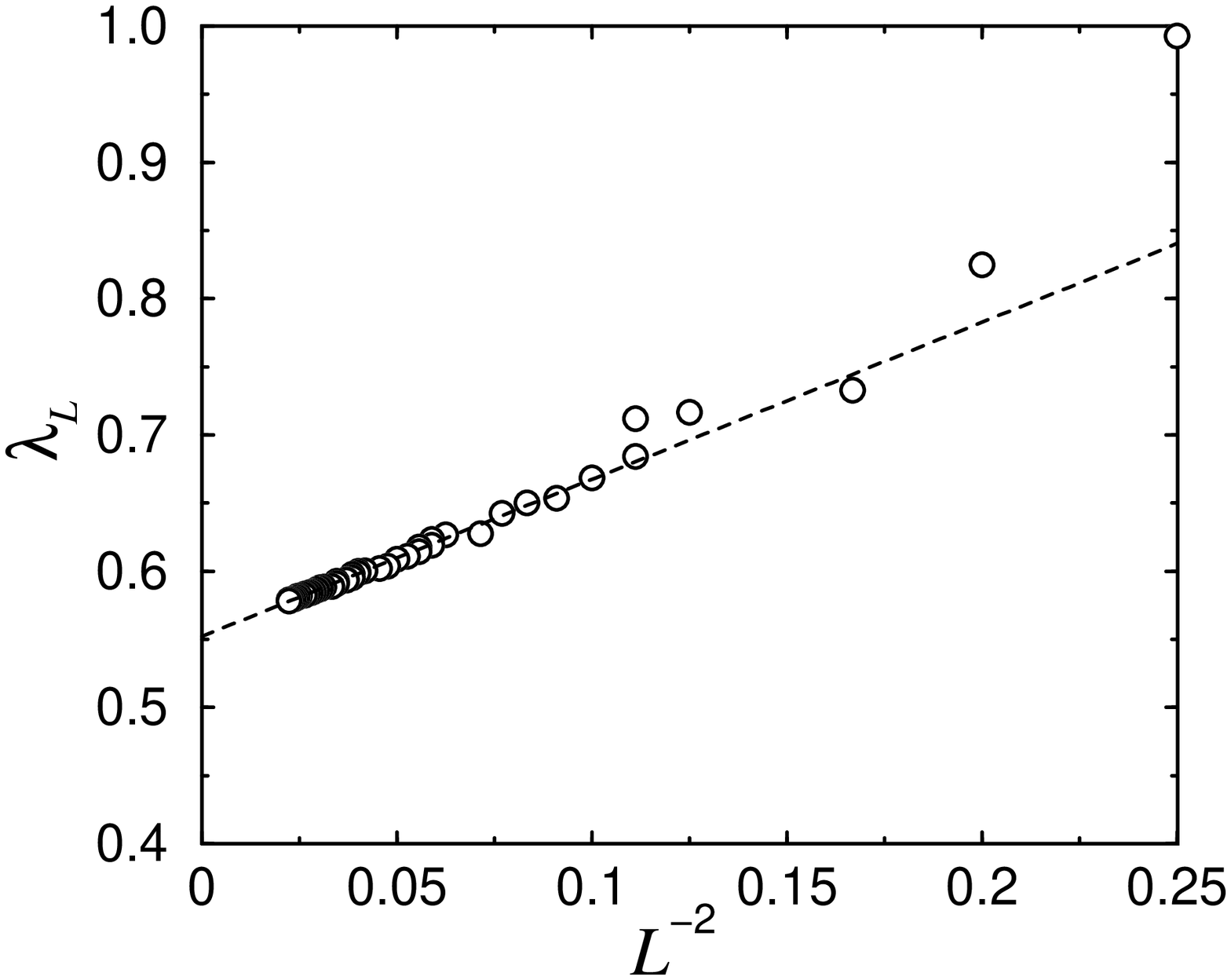}}
\caption{The SUB2 termination point as a function of $L^{-2}$ defined in Eq.~(\ref{lambda})
in 3 space dimensions, using the CCM operatorial form.}
\label{fig:lam_end_3d}
\end{figure}

As we have noted earlier, the
convergence to the limit point is fast in terms of the number $m$ of basis
functions required, probably indicating that the CCM is a good way to
evaluate the ground-state properties. The number $n$ of pairwise correlation
 components is the
major factor in the convergence, as one would expect at a phase
transition where the correlation length becomes infinite.
In order to extrapolate the SUB2 critical value of $\lambda$ corresponding to 
the termination point, we perform a fit of the form
\begin{equation}
\lambda_L = \lambda_\infty +c/L^2 ,
\label{lambda}
\end{equation}
where $L$ is the 
distance between the correlated pairs which are furthest apart in the 
SUB2-$n$ calculation for a given value of the truncation index $n$. 
This asymptotic form is heuristic rather than having any theoretical 
basis. We start this
extrapolation at $L=3$, and make a standard linear fit to the numerical
results, as indicated by the dashed lines in 
Figs.~\ref{fig:lam_end_1d}--\ref{fig:lam_end_3d} for the operatorial
cases. We contrast the results
of the functional and operatorial approaches in Table~\ref{tab:compare}.


To our surprise, and somewhat to our dismay, we find that both the 
functional and operatorial methods predict a termination point, possibly 
indicative of a phase transition, in one dimension. This probably 
shows a defect in our approximation scheme, since no 
transition  of this type is expected for such a one-dimensional system, and the exact results by 
Polyakov and Wiegmann~\cite{PW83} for the Euclidean theory also show 
no phase transition.
The positions of the calculated  termination
points are surprisingly similar for the two approaches in all numbers of spatial
 dimensions,
 however.
We also believe that both calculations have converged, so there is no 
obvious numerical explanation for the small differences.

Of course we have not yet addressed the important issue of
 whether our results indicating
a phase transition at finite $\lambda$  survive 
in higher-order approximations.  It will be interesting to pursue this matter,
especially for the 1D case (where such a more sophisticated calculation 
to incorporate correlations between triplets or other $n$-body clusters with
$n>2$ may be more feasible 
than in higher spatial dimensionality).

\begin{table}
\begin{center}
\begin{tabular}{|c|c|c|}
\hline
dimension  & \multicolumn{2}{c|}{$\lambda_\infty$} \\ \cline{2-3}
           & functional & operatorial\\
\hline 
1 &  1.927 & 2.088\\
2 &  0.827 &  0.887\\
3 &  0.536  &  0.552\\
\hline
\end{tabular}
\end{center}
\caption{The extrapolated CCM SUB2 termination points for the functional and operatorial
 methods. 
 }
\label{tab:compare}
\end{table}

\subsection{Excitations}

\begin{figure}[htb]
\centerline{\includegraphics[height=7cm]{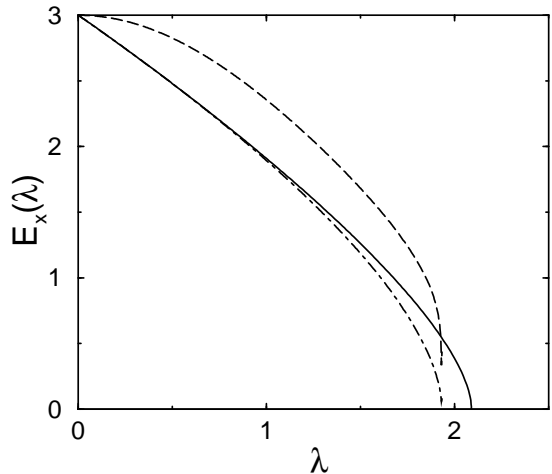}}
\caption{The excitation energy as a function of $\lambda$
in 1 space dimension in the CCM SUB2 approximations.
 The solid line is the RPA result from the 
operatorial approach; the dashed line the values obtained using the 
Feynman technique
 in the functional approach, and the dashed-dotted line the 
corresponding RPA result using the functional approach.}
\label{fig:Ex_1d}
\end{figure}
\begin{figure}[htb]
\centerline{\includegraphics[height=7cm]{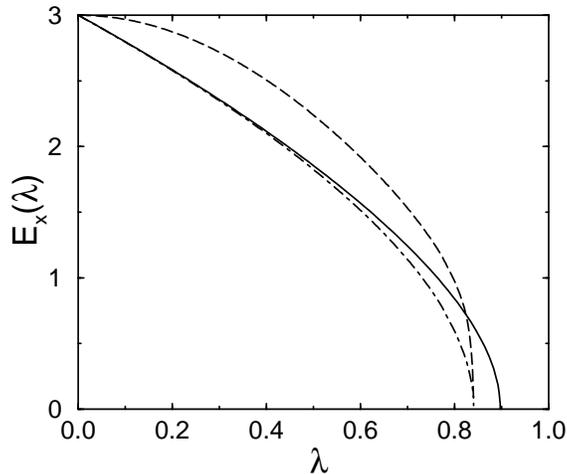}}
\caption{The excitation energy as a function of $\lambda$
in 2 space dimensions in the CCM SUB2 approximations. 
For the meaning of the lines see Fig.\ 
\protect{\ref{fig:Ex_1d}}.}
\label{fig:Ex_2d}
\end{figure}
\begin{figure}[htb]
\centerline{\includegraphics[height=7cm]{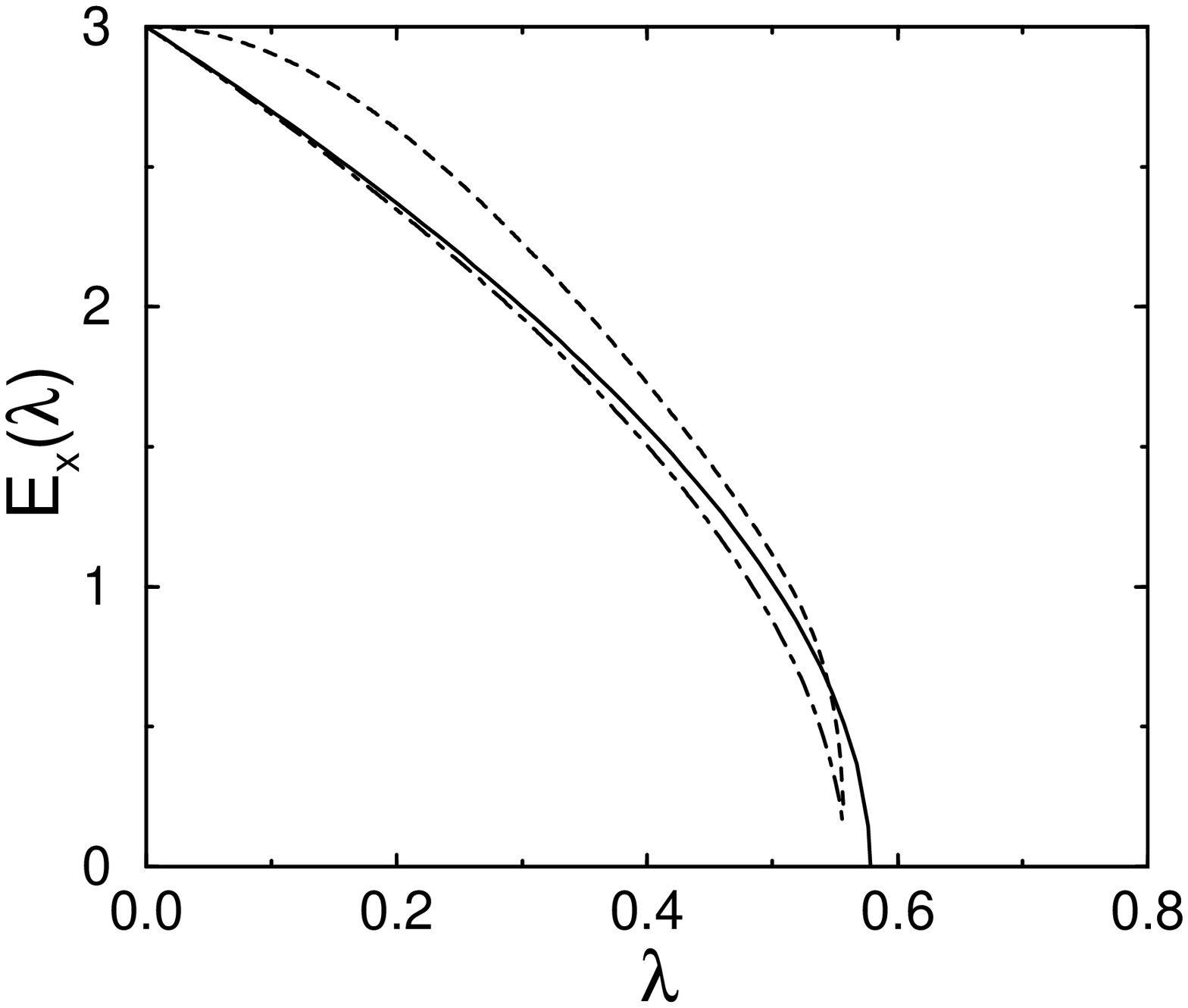}}
\caption{The excitation energy as a function of $\lambda$
in 3 space dimensions in the CCM SUB2-$n$-$m$ approximations at
the highest practical truncation indices $n$ and $m$.  Note
that (for three dimensions)
convergence in the index $n$ near the termination point is not fully attained.
For the meaning of the lines see Fig.\ 
\protect{\ref{fig:Ex_1d}}.}
\label{fig:Ex_3d}
\end{figure}

Excitation energies were evaluated by the linear response technique,
for both the functional and operatorial CCM methods, as well as by 
Feynman's technique for the functional method only.
The solutions for 1,2, and 3 dimensions are given in 
Figs.~\ref{fig:Ex_1d}--\ref{fig:Ex_3d}. In all three cases we find 
that the excitation energy obtained from the Feynman technique lies above 
that obtained from the corresponding linear
response (RPA) technique. As we approach the termination points
the excitation energies approach zero rapidly. 
 However, since the numerical calculations become unstable
at the the termination point, Feynman's technique only
gives a very strong indication that at this point
the system becomes gapless. 

The RPA linear response methods, where we expand around the 
ground-state solution,
give slightly lower values for the excitation energy. In these calculations
the excitation can be tracked all the way to the termination point, where the
system becomes gapless. There is some
doubt whether the extremely small difference between the point where 
the solution 
terminates and where the RPA frequency goes through zero is due to 
numerical artefacts or has some physical significance, indicating
that only the full SUB2 calculation will be exactly gapless at the
termination point. This might, a priori, be expected since
only the full SUB2 calculation contains
the arbitrarily long-range correlations needed precisely at a 
phase transition point.

So what is the interpretation of the phase transitions that we have seen? 
We have 
investigated the eigenvectors for the mode that becomes (almost) 
gapless, but unfortunately the structure of these vectors, which tell 
us about the excitation operators to the new vacuum, do not seem to 
show any discernable regularity. This is probably related to the inability of the 
present method to describe the state beyond the putative phase 
transition.

\begin{figure}[htb]
\centerline{\includegraphics[height=7cm]{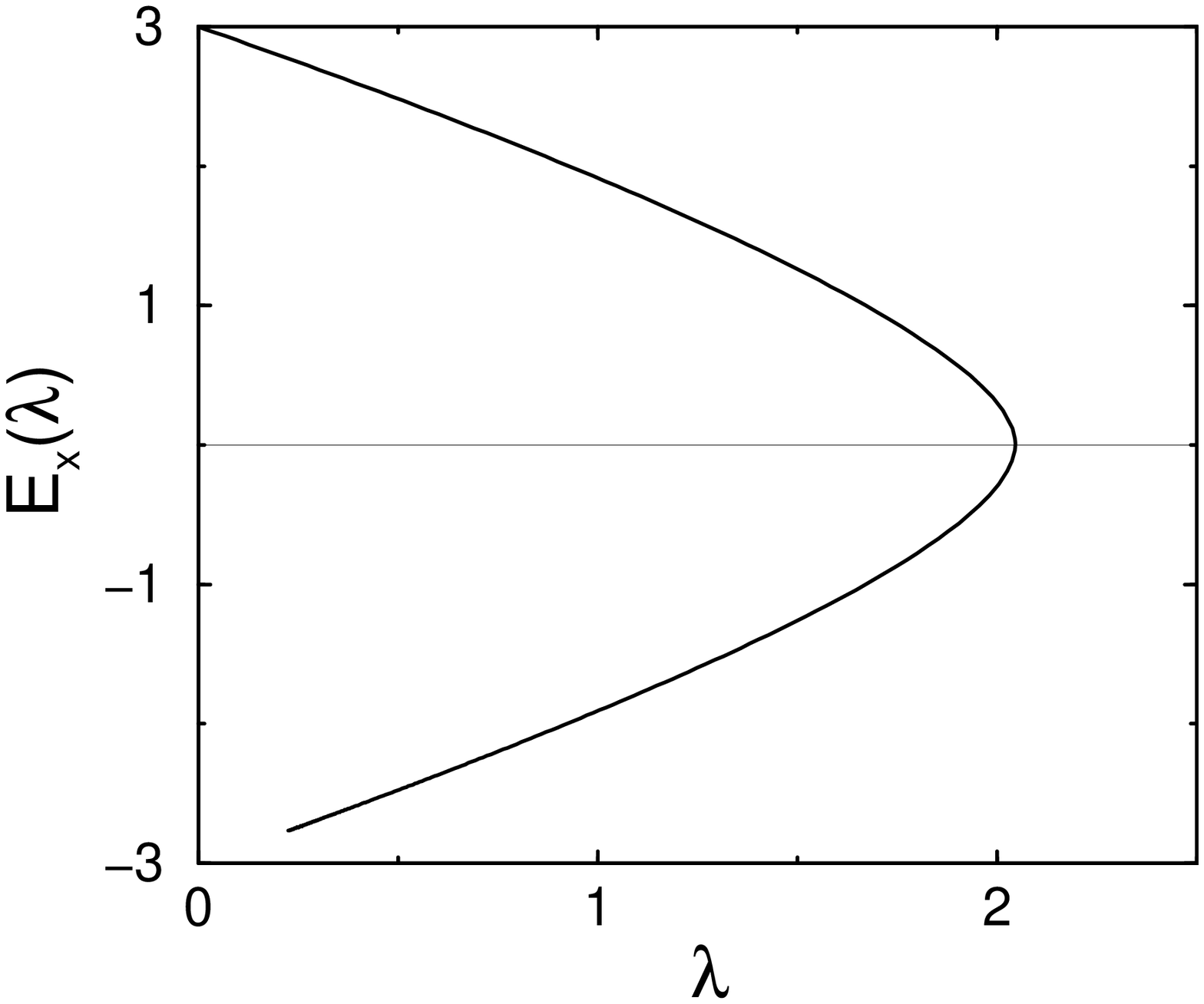}}
\caption{The excitation energy as a function of $\lambda$
in 1 space dimension using the operatorial form of the CCM in SUB2
approximation, showing both the physical and
unphysical branches.}
\label{fig:back}
\end{figure}

The study of the linear response also allows us to display most vividly
the occurrence of two branches in the energy spectrum. In Fig.~\ref{fig:back}
we show the lowest excited state, where we follow the solution of
the 1D SUB2-$n$ equations to the ``termination point'' and beyond. At 
this point the solution turns around, and we find a second solution for
each value of $\lambda$ where the ground-state energy is extremely
close to the one from the physical solution (so close that it is impossible
to distinguish the differences in the figures).
We can follow this solution 
until we return close to the point $\lambda=0$, where some of the CCM coefficients
diverge. The excitation energy in this second branch is negative, a not
very desirable situation. However, the fact that the second branch does not 
return to the known physical ground state at $\lambda=0$, and that 
the excitation energies are
negative leads us to denote this branch as being unphysical, an artefact of
the CCM.

\subsection{Comparison with other work}

There exist only two sets of results with which we can easily and
 directly compare our
results, namely the calculation by Chin \cite{Shi97} using the CBF
technique, and a calculation using the strong-coupling expansion by
Sobelman \cite{HKS79}.

Sobelman has calculated the strong-coupling expansion in a Hamiltonian
framework for a general $O(N)$ system in  an arbitrary number of spatial
 dimensions up to fourth
order. If we use his results to calculate the coupling constant for which
the mass gap disappears we find systematically larger values for the (2+1)-
and the (3+1)-dimensional models, by a factor of about 3, than obtained 
by our results.
In the third-order expansion there is a vanishing mass gap 
for (1+1) dimensions at the critical value $\lambda=8$, 
but this criticality disappears in the fourth order. There remains a dip in the
excitation energy around $\lambda = 8 $ where it crossed the zero axis
before. From the exact results this seems to be the correct behaviour.
However, the large change between third and fourth orders suggests that
the expansion has not converged for the (1+1) model in this regime.
For (2+1) and (3+1) dimensions the positions of the vanishing mass gap
change somewhat, as one goes from third order to fourth order 
(and see Table~\ref{tab:sobelman}).  The convergence properties of the series are,
however, completely unknown.

\begin{table}
\begin{center}
\begin{tabular}{|c|c|c|}
\hline
dimension  & \multicolumn{2}{c|}{$\lambda_\infty$} \\ \cline{2-3}
           & third order & fourth order\\
\hline
1 &  8. & -- \\
2 &  2.624 &  2.464 \\
3 &  1.637  &  1.492 \\
\hline
\end{tabular}
\end{center}
\caption{The coupling constants for which the excitation energies
vanishes, determined from Sobelman's strong coupling expansion.}
\label{tab:sobelman}
\end{table}

Because of a different definition of the kinetic energy operator,  the
value obtained by Chin \cite{Shi97} for the critical
coupling constant needs to be multiplied by a factor of four to agree with our
definition of the model Hamiltonian given by Eq.~(\ref{hamil}). His results are
for three dimensions only and he finds a critical coupling constant of
$\lambda \approx 0.96$  (after scaling) compared to our results $\lambda_\infty = 0.54$  
from the SUB2 functional form and 
$\lambda_\infty = 0.55$ from the SUB2 operatorial operatorial.
 However, more strikingly, he finds
a first-order phase transition, while all our results indicate a second-order
phase transition.

\section{Conclusions and outlook}\label{sec:Conc}

Our results are an interesting  albeit still
limited approach to a nontrivial field theory which exhibits a phase transition.
 They show that we 
can construct two different CCM-like approaches for field theories, 
which are amenable to different sets of approximations. For similar 
truncations the two methods produce very similar results.

In one space dimension we would not have expected a termination point or a phase
transition from known exact results. The fact 
that in our approximations the termination point occurs at a large value of $\lambda$ 
shows that it is probably not stable against the correlations missing from our SUB2 calculations.
The corresponding stability of our results in two and three 
spatial dimensions is certainly also of interest.
The current calculations cannot address this issue, since they would have to 
be supplemented by complicated higher-order calculations in order to make 
stringent statements. This further raises the question whether one  should not 
aim to apply the extended coupled cluster method (ECCM)
\cite{Arp83}
to the present problem. The ECCM generally has been shown to be more powerful 
than the normal (NCCM) version of the CCM used here
in describing states {\it on both sides of a phase transition}, and might 
give us an idea of its nature. We believe that this might be an 
interesting problem for a future study.

Finally it would be appealing to look at real gauge field theories 
such as compact QED, or even QCD, in the light of our results. We 
plan to see whether the lessons we have learned in this work have any 
relevance for the methods of Refs. \cite{BBD96,BBD97}. We believe 
that armed with our arsenal of techniques we should be ready to tackle 
such problems, and even  be able to proceed towards the
inclusion of dynamical quarks 
in such calculations

\vspace{5mm}
\noindent
{\Large \bf Acknowledgement } \\
\vspace{3mm}

\noindent
We acknowledge support through a research grant from the Engineering and 
Physical Sciences Research Council (EPSRC) of Great Britain.

\appendix
\section{Quantisation} \label{app:Quant}
The systematic quantisation of the non-linear sigma model is highly nontrivial.
Using the components of the unit vector $\fuvc n$ as degrees of freedom,
the constraint among them ($\fuvc{n}^2 = 1$) leads to a second constraint
from the time independence of the first constraint. The Dirac brackets,
which restrict the quantised operators to the physical subspace, 
are not particularly difficult to deal with \cite{Cebula}, but they lead to a form that
does not readily lend itself to the kind of techniques we would like to
apply. One can also avoid using constraints by using an exponential
parametrisation, $U=e^{i \vec \tau \cdot \vec \theta}$. However, this leads to
even more complicated expressions.

We choose to use the unit vectors $\fuvc n$ as degrees of freedom, but
avoid the use of constraints by using an Euler angle parametrisation
of these vectors,
\be
\fuvc  n = (\sin\theta\sin\phi\sin\chi,
\sin\theta\sin\phi\cos\chi,
\sin\theta\cos\phi,
\cos\theta),
\end{equation}
with $0<\theta<\pi$,  $0<\phi <\pi$, and  $0<\chi < 2 \pi$.
 Since the Laplacian is separable in spherical
coordinates, we find that the kinetic energy $K$  at a single lattice
point is just the half of the angular part of the Laplacian,
 which is proportional
to the Casimir invariant of the relevant $O(N)$ group.
The kinetic energy can be derived from the generalized angular momentum operators
\be
L_{ij} = -i(n_i\partial_{n_j} - n_j\partial_{n_i} ),
\end{equation}
where $i,j = 1,2,\cdots, N$.
Specifically, it may
can be expressed as (the $O(N)$ Casimir invariant or) the sum-of-squares 
of the  angular momentum operators,
\be
2 K= \sum^N_{i<j} L_{ij}^2
 = -\partial^2_\theta - (N-2) \cot \theta \partial_\theta -
 \frac{1 }{ \sin^2 \theta} \partial_\phi^2 - \frac{ (N-3) \cot \phi }{
 \sin^2 \theta} \partial_\phi - \frac{1 }{ \sin^2 \theta \sin^2 \phi } 
\partial^2_\chi + \cdots
\end{equation}
In many cases we shall only need the $\theta$-dependent part,
which can be rewritten in terms of $\cos \theta$ as
\be
-(1- \cos^2 \theta) \partial^2_{\cos \theta} + (N-1) \cos \theta \partial_{\cos \theta}.
\end{equation}

\section{Gegenbauer polynomials} \label{app:Gegenbauer}

The Gegenbauer polynomials, $C^\alpha_n(x)$, are orthogonal polynomials 
on the domain $[-1,1]$
with respect to the weight $(1-x^2)^{\alpha -1/2}$. The integration
measure for the azimuthal angle on an $N$-dimensional sphere is 
$\sin^{N-2} \theta$, which means that orthogonal functions
on the $N$-sphere are $C^\alpha_n(\cos \theta)$ with $\alpha = (N-2)/2$. 
The Gegenbauer polynomials are eigenfunctions of the kinetic energy 
operator
\be
-[(1-x^2) \partial_x^2 - (N-1) x \partial_x ] C^{(N-2)/2}_n(x) = n(n+N-2) C^{(N-2)/2}_n(x).
\end{equation}
For $N<4$ there is a second, non-singular set of eigenfunctions of the relative
angle 
\be
-[(1-x^2) \partial_x^2 - (N-1) x \partial_x ] (1-x^2)^{(3-N)/2} C^{(4-N)/2}_n(x) = (n+1)(n-N+3) C^{(4-N)/2}_n(x),
\end{equation}
which coincides with the first set of solutions for $N=3$, 
and leads to a set of sine functions for $N=2$. For a general discussion of
the properties of the Gegenbauer polynomials we refer to \cite{MOS}.
Here we shall just summarise a few key points.

One relation that we shall use frequently in the paper is related to the 
addition theorem for Gegenbauer polynomials. This occurs in the integration
over intermediate points in a chain of functions linking sets of lattice
points. Only products of Gegenbauer polynomials of the same order contribute
to the integral. This holds for arbitrary dimensions, and allows us to
extend our methods to any $O(N)$ theory. Using the addition theorem for 
Gegenbauer polynomials
\[
C_n^\alpha(\cos \theta_1 \cos \theta_2 + \sin \theta_1 \sin \theta_2 \cos \phi)=
\]
\[
\frac{\Gamma(2 \alpha -1) }{ \Gamma^2(\alpha)} \sum_{m=0}^n
\frac{2 ^{2m} \Gamma (n-m+1) (\Gamma(\alpha +m))^2 }{ \Gamma(n + 2 \alpha +m)}
(2 \alpha + 2 m-1)  
\]
\begin{equation}
\times (\sin \theta_1)^m (\sin \theta_2)^m C_{n-m}^{\alpha+m}
(\cos \theta_1) C_{n-m}^{\alpha+m}(\cos \theta_2) C^{\alpha-\frac{1 }{2}}_m
(\cos \phi)  ,
\end{equation}
and orthogonality of the Gegenbauer
polynomials it follows that
\begin{equation}
\int \frac{d \mu (n_{\vec{i}}) }{ \Omega_N}  
C^\alpha_k(\hat n_{\vec{i}} \cdot \hat n_{\vec{j}})
           C^\alpha_l(\hat n_{\vec{i}} \cdot \hat n_{\vec{p}})
= \frac{\sqrt{\pi} \Gamma(1+\alpha)4^{1-\alpha} \Gamma(2 \alpha -1) }{ \Gamma^2(\alpha)(\alpha+k) \Gamma(\alpha + 1/2)}
\delta_{kl} C^\alpha_k(\hat n_{\vec{j}} \cdot \hat n_{\vec{p}}).
\end{equation}
When integrating over closed loops this general property implies
that only ``local terms'' contribute, i.e., products of 
Gegenbauer polynomials of identical order.
For $N=4$ where $\alpha=1$ this reduces to
\begin{equation}
\int \frac{d \mu (n_i) }{ \Omega_4}  U_k(\hat n_i \cdot \hat n_j)
                                     U_l(\hat n_i \cdot \hat n_p)
= \frac{ \delta_{kl} U_k(\hat n_j \cdot \hat n_p) }{ 1 + k},
\end{equation}
where $C^1_n(\cos \theta) \equiv U_n(\cos \theta)$ is the Chebyshev 
polynomial of the second kind.

The fact that
the derivative of a polynomial is again a 
polynomial
 allows us  to express the derivative of a Gegenbauer polynomial 
as a weighted sum of Gegenbauer polynomials
\be
\partial_x C^\alpha_n(x) = \sum_{i=0}^{[n/2]} 2(n-1+\alpha -2 i) 
C^\alpha_{n-2i-1}(x).
\end{equation}
We shall apply this transformation for the case where $n\neq 0$ (we exclude
a constant part from our correlation functions). In that case we can define
a matrix $T$ by 
\be
T_{nm} =\sum_{i=0}^{[n/2]} 2(m+\alpha) \delta_{m,n-2i-1},
\label{Gegder}
\end{equation}
which in the subspace of interest is invertible. This linear operator $T$ is
used in Eq.~(\ref{funcRPA}) in the RPA for the collective
excited states.


\begin{thebibliography}{50}
\bibitem{DGH92} J.F. Donoghue, E. Golowich, and B.R. Holstein, {\it Dynamics of
                the Standard Model}, Cambridge Press, Cambridge (1992).
\bibitem{GOR} M. Gell-Mann, R. Oakes, and B. Renner, Phys. Rev. {\bf 175}
(1968) 2195.
\bibitem{BrownRho} G.E. Brown and M. Rho, Phys. Rev. Lett. {\bf 66} (1991)
2720.
\bibitem{BW94} G. Bunatian and J. Wambach, Phys. Lett. {\bf B336} (1994) 290.
\bibitem{BchPt} M.A. Nowak, M. Rho, and I. Zahed, {\it Chiral Nuclear Dynamics},
                World Scientific, Singapore (1996).            
\bibitem{PW83} A. Polyakov and P.B. Wiegmann, Phys. Lett. {\bf B131} (1983) 121.
\bibitem{PW84} R.D. Pizarski and F. Wilczek, Phys. Rev. D {\bf 29} (1984) 338.
\bibitem{BZG76} E. Br\'ezin, J. Zinn-Justin, and J.C. Le Guillou, Phys. Rev.
                D {\bf 14} (1976) 2615.
\bibitem{BC96} P. Butera and M. Comi, Phys. Rev. B {\bf 54} (1996) 15828; 
          M. L\"uscher and P. Weisz, Nucl. Phys. { \bf B300}[FS22] (1988) 325 (and
                references contained therein);
          J. Kogut. M. Snow, and M. Stone, Nucl. Phys. {\bf B200}[FS4] (1982) 211.
\bibitem{Sta68} H.E. Stanley, Phys. Rev. {\bf 176} (1968) 718.
\bibitem{Shi97} S.A. Chin, {\it Advances in Quantum Many-Body Theory}, Vol.
{\bf 1}, (eds.) D. Neilson and R.F. Bishop, World Scientific, Singapore (1998), 
to be published.
\bibitem{BishopChim} R.F. Bishop, Theor. Chim. Acta {\bf 80} (1991) 95.
\bibitem{BBD96} S.J. Baker, R.F. Bishop, and N.J. Davidson, Phys. Rev. D {\bf 53}
                (1996) 2610.
\bibitem{BBD97} S.J. Baker, R.F. Bishop, and N.J. Davidson, Nucl. Phys. B
               (Proc. Supp.) {\bf 53} (1997) 834.
\bibitem{Arp97}  J. Arponen, Phys. Rev. A {\bf 55} (1997) 2686.
\bibitem{spin} R.F. Bishop, J.B. Parkinson and Y. Xian, Phys. Rev. 
B {\bf 44} (1991) 9425. 
\bibitem{AS64} M. Abramowitz and I. Stegun, {\it Handbook of Mathematical Functions}, National Bureau of Standards, Applied Mathematics Series 55, 
               Washington, D.C. (1964).
\bibitem{Rheinboldt} W.C. Rheinboldt, 
{\em Methods for Solving Systems of Nonlinear Equations}, SIAM, Philadelphia, (1987);
{\em Numerical analysis of parametrized nonlinear equations}, Wiley, New York (1986).
\bibitem{Pitcon} W.C. Rheinboldt and J. Burkardt, ACM TOMS  {\bf 9} (1983) 236.
\bibitem{MW66} N.D. Mermin and H. Wagner, Phys. Lett. {\bf 17} (1966) 1133.
\bibitem{HKS79} C.J. Hamer, J.B. Kogut, and L. Susskind, Phys. Rev. D {\bf 19}
                (1979) 3091; G.E. Sobelman, Phys. Rev. B {\bf 24} (1981) 1493.
\bibitem{Arp83} J. Arponen, Ann. Phys. (N.Y.) {\bf 151} (1983) 311.
\bibitem{Cebula} D.P. Cebula, A. Klein, and N.R. Walet, Phys. Rev. D {\bf 47} (1993) 2113.
\bibitem{MOS}W. Magnus, F. Oberhettinger, and R.P. Soni, {\it Formulas and 
Theorems for Special Functions of Mathematical Physics}, 
             Springer-Verlag, Berlin (1966).
\end{thebibliography}
\end{document}